\documentclass{article}

\usepackage{multirow}% http://ctan.org/pkg/multirow (for KK matrix)
\usepackage{PRIMEarxiv}
\usepackage{authblk}
\usepackage{amsmath}
\usepackage{cancel}
\usepackage{makecell}
\usepackage[utf8]{inputenc} % allow utf-8 input
\usepackage[T1]{fontenc}    % use 8-bit T1 fonts
\usepackage{hyperref}       % hyperlinks
\usepackage{url}            % simple URL typesetting
\usepackage{booktabs}       % professional-quality tables
\usepackage{amsfonts}       % blackboard math symbols
\usepackage{nicefrac}       % compact symbols for 1/2, etc.
\usepackage{microtype}      % microtypography
\usepackage{lipsum}
\usepackage{fancyhdr}       % header
\usepackage{graphicx}       % graphics
\usepackage{caption}
\usepackage{subcaption}
\usepackage[titletoc]{appendix}
\usepackage{titlesec}
\usepackage{caption} 
\graphicspath{{media/}}     % organize your images and other figures under media/ folder

\newcommand{\conv}{\text{conv}} 
\newcommand{\rank}{\text{rank}} 

\newcommand{\argmax}{\text{argmax}}

\newcommand{\K} {\mathbf{K'}} % Surface current
\newcommand{\KK} {\K\cdot\nabla\K} % Surface current
\newcommand{\N} {\mathbf{N}} % A non-unit surface norm
\newcommand{\n} {\mathbf{n}} % A unit surface norm
\newcommand{\B} {\mathbf{B}} % The magnetic field
\newcommand{\rr} {\mathbf{r}} % The position
\newcommand{\vv} {\mathbf{v}} % Arbitrary vector field for illustrative purposes
\newcommand{\Sset} {\mathbb{S}} % Set of positice semidefinite matrices.
\newcommand{\Rset} {\mathbb{R}} % The position
\newcommand{\Dset} {\mathbb{D}} % The domain
\newcommand{\Hset} {\mathbb{H}} % The domain 's convex hull

\title{Global stellarator coil optimization with quadratic constraints and objectives}

\author[1,2]{Lanke Fu}
\author[3]{Elizabeth J. Paul}
\author[4]{Alan A. Kaptanoglu}
\author[2]{Amitava Bhattacharjee}
\affil[1]{Princeton Plasma Physics Laboratory, Princeton, NJ}
\affil[2]{Department of Astrophysical Sciences, Princeton University, Princeton, NJ}
\affil[3]{Department of Applied Physics, Columbia University, New York, NY}
\affil[4]{Courant Institute of Mathematical Sciences, New York University, New York, NY}

\begin{document}
\maketitle

\begin{abstract}
Most present stellarator designs are produced by costly two-stage optimization: the first for an optimized equilibrium, and the second for a coil design reproducing its magnetic configuration. Few proxies for coil complexity and forces exist at the equilibrium stage. Rapid initial state finding for both stages is a topic of active research. Most present convex coil optimization codes use the least square winding surface method by Merkel (NESCOIL), with recent improvement in conditioning, regularization , sparsity and physics objectives. While elegant, the method is limited to modeling the norms of linear functions in coil current. We present QUADCOIL, a fast, global coil optimization method that targets combinations of linear and quadratic functions of the current. It can directly constrain and/or minimize a wide range of physics objectives unavailable in NESCOIL and REGCOIL, including the Lorentz force, magnetic energy, curvature, field-current alignment, and the maximum density of a dipole array. QUADCOIL requires no initial guess and runs nearly $10^2\times$ faster than filament optimization. Integrating it in the equilibrium optimization stage can potentially exclude equilibria with difficult-to-design coils, without significantly increasing the computation time per iteration. QUADCOIL finds the exact, global minimum in a large parameter space when possible, and otherwise finds a well-performing approximate global minimum. It supports most regularization techniques developed for NESCOIL and REGCOIL. We demonstrate QUADCOIL's effectiveness in coil topology control, minimizing non-convex penalties, and predicting filament coil complexity with three numerical examples.
\end{abstract}
% minimizing total field error $\int_{plasma}d^2a|B_{normal}|^2$ and , $max\|\K\cdot\nabla \K\|_\infty$, proxy for the maximum curvature of a surface current $\K$.

% keywords can be removed

\section{Introduction}\label{section:intro}

Stellarators are attractive three-dimensional (3D) fusion devices that promise steady-state, disruption-free operation. Unlike tokamaks, stellarators generate rotational transform by external coils rather than a plasma current. A stellarator reactor's coil system determines dominantly its cost and performance~\cite{strykowsky_engineering_2009}. 

Traditionally, stellarator designs are produced by two-stage optimization: the first stage for an equilibrium with MHD stability and good particle confinement, and the second stage for a coil set that accurately reproduces the equilibrium magnetic field. Both stages are high-dimensional in parameter space, nonlinear, and ill-posed optimization problems. The coil stage, specifically, is an inverse Biot-Savart problem that does not have a unique solution.

Presently, there are two main methods for coil optimization: filament and winding surface. The filament method treats coils as discrete space curves. Examples of filament codes include COILOPT \cite{strickler_designing_2002}, FOCUS \cite{zhu_new_2018}, SIMSOPT \cite{landreman_simsopt:_2021}, and DESC \cite{dudt_desc_2020}. This method has been shown to generate realistic coil sets. However, because it formulates coil optimization as a non-convex optimization problem, the filament method suffers from many common issues in high-dimensional, multi-objective non-convex optimization. Non-convex optimization is generally NP-hard. Good convergence typically requires careful manual tuning on the initial state and penalty weights. Moreover, in practice, several filament sub-problems are solved sequentially, starting with low-resolution solutions that are used to initialize optimization at increasing resolutions. The coil number and topology in the filament method are also fixed during the optimization. Due to these disadvantages, designing a coil set using the filament method is often computationally costly and labor-intensive.

The winding surface method approximates a coil set as a sheet current $\K$ on a prescribed winding surface, then cuts the sheet current into discrete coils by choosing a set of current paths and giving each path an appropriate current. Examples of such codes include NESCOIL \cite{merkel_solution_1987}, ONSET \cite{drevlak_coil_1998},  and REGCOIL \cite{landreman_improved_2017}. Compared to the filament method, this method is less realistic, but it formulates coil optimization as a convex optimization problem. Therefore, the winding surface method has the theoretical complexity of $\mathcal{O}(n_\text{sample}n_{\text{dof}}^2)$, where $n_\text{sample}$ is the number of sample points in the field error calculation, and $n_{\text{dof}}$ is the number of degrees of freedom of that encodes $\K$. Although the method requires an arbitrary choice of winding surface, it always converges to the global optimum regardless of the surface choice. In practice, the filament method and winding surface method are often used in complement. The filament method produces the final coil set, whereas the winding surface method provides initial states for the filament method \cite{drevlak_automated_1998}.

The winding surface method is naturally suitable as a proxy for coil complexity at the equilibrium stage. The method has been used to study the impact of plasma parameters on coil complexity \cite{kappel_magnetic_2024}. However, despite recent improvements, \cite{boozer_optimization_2000,landreman_improved_2017,pomphrey_innovations_2001,kaptanoglu_topology_2024}, its formulation greatly limits the choice of objectives and constraints. All existing winding surface methods are limited to minimizing the L-1 or L-2 norms of linear functions of $\K$. This greatly limits the choice of physics and engineering metrics. All existing winding surface methods are limited to performing unconstrained optimization. This limits its control over the topology of $\K$. As a result, the sheet current may contain complex features that make coil-cutting difficult.

We present QUADCOIL, a reformulation of the winding surface method that enables global coil optimization using a non-convex quadratic objective under non-convex quadratic constraints. QUADCOIL can directly target a wide range of new metrics. It offers greater control over the topology of $\K$, allowing automated coil-cutting with no human intervention. QUADCOIL is fast and global. Similar to existing winding surface methods, it requires no initial guess and minimal parameter tuning. This makes it suitable for initial state generation or as a coil complexity metric in the equilibrium optimization loop. Although it is tuned for fusion applications, QUADCOIL is fundamentally an inverse Biot Savart code and can be used for other magnetostatic optimization problems, such as accelerator magnet design.

Notably, in recent years, single-stage stellarator optimization has risen in popularity. Single-stage optimization simultaneously optimizes the stellarator equilibrium and coil set to find a compromise between physics and coil complexity requirements \cite{giuliani_single-stage_2022}\cite{conlin_free_2022}. Alternatively, the approach can also design stellarators that are robust to coil manufacturing errors. While the single-stage approach has successfully produced a large number of optimized equilibria with quasisymmetry (QS), its parameter space may be needlessly large. Because one equilibrium can be reproduced by many different coil configurations, similar equilibria can potentially appear in multiple, distant regions of the large parameter space. This may increase the number of local minima compared to those in either stage of the two-stage approach, which makes the optimization process more costly and harder to converge. In comparison, QUADCOIL connects the coil and equilibrium stage during a two-stage optimization as a coil complexity metric. We believe our approach can potentially balance plasma performance and coil complexity, similar to a single-stage optimization, but without introducing additional degrees of freedom.

Our paper is organized as follows. Section \ref{section:theory:NESCOIL} introduces the theory behind existing winding surface methods and surveys recent developments. Section \ref{section:theory:objectives} discusses QUADCOIL's formulation and motivations. Section \ref{section:numerical:topology} - \ref{section:numerical:parameter space} demonstrate QUADCOIL's effectiveness in coil topology control, minimizing non-convex penalty, and predicting filament coil complexity with three numerical examples.

\section{Theory}

\subsection{Winding surface method}
\label{section:theory:NESCOIL}

In this section, we will briefly introduce the formulation shared by existing winding surface methods, their advantages, and their limitations.

Stellarator coil optimization, at its core, is an inverse Biot-Savart problem. Typically, this problem is non-convex, due to the $|\rr-\rr'|^{-3}$ non-linearity in the Biot-Savart Law (In this paper, primed letters will denote coil-related quantities, and unprimed letters will denote plasma-related quantities)
\begin{equation}
    \B(\mathbf r) = \int d^2a' \frac{\mathbf{I}'\times(\rr-\rr')}{|\rr-\rr'|^3}.
\end{equation}
Here, $\mathbf{I}'$ is the coil current vector, and $\rr$ and $\rr'$ are locations at the plasma boundary and the coils. The winding surface method circumvents this non-linearity by fixing the location of the current. Because $\K$ is divergence-free, we can represent it with the gradient of a current potential $\Phi'$ \cite{merkel_solution_1987}:
\begin{equation}
    \nabla \cdot \K=0 \Leftrightarrow \K=\hat \n'\times\nabla\Phi',
\end{equation}
where $\hat \n'$ is the unit normal of the winding surface. Generally, the current potential $\Phi'$ consists of three components:
\begin{equation}
    \Phi' = \Phi'_{\text{sv}}(\zeta', \theta')+\frac{G\zeta'}{2\pi}+\frac{I\theta'}{2\pi}.
\end{equation}
Here, $\zeta', \theta'$ are arbitrary toroidal and poloidal angles on the winding surface and $\Phi'_{\text{sv}}$ is the single-valued component of $\Phi'$, the degrees of freedom we optimize. $G, I$ are the net toroidal and poloidal currents. The value of $I$ is determined by the toroidal flux of the equilibrium, whereas the value $G$ is a free parameter that determines whether $\K$ forms primarily poloidal or helical coils. For brevity, we will abbreviate $\Phi'_{\text{sv}}$ as $\Phi'$ in the rest of this paper.

To visualize $\Phi'$, the current $\K$ always flows along contours of equal $\Phi'$. The ratio between $I$ and $G$ determines the helicity of the net current. Alternatively, the current potential $\Phi'$ has the unit of surface magnetic dipole moment density, and its single-valued component, $\Phi'_\text{sv}$, can be considered as a sheet of dipoles oriented perpendicular to the winding surface. Therefore, the winding surface approach is suitable for modeling a variety of coil types, including poloidal coils, helical coils, windowpane coils, and dipole arrays. Using multiple winding surfaces with different objectives, constraints and net currents, it may also be possible to model a combination of different coil types.

Since the Biot-Savart Law is linear in $\K$, we can minimize $f_B$, the squared flux at the plasma boundary, as a linear least square problem,
\begin{gather}\label{equation:cp:nescoil} 
    \min_{\Phi'}f_B(\Phi') = \min_{\Phi'}\|A_B\Phi'+b_B\|_2,\\
    f_B\equiv\int_{\text{plasma}}d^2 a |\B_\text{coil}\cdot\hat \n - B_T|^2.
\end{gather}
Here, $\hat \n$ is the surface normal of the plasma surface, $B_\text{coil}$ is the magnetic field produced by the stellarator coils, and $B_T$ is the target normal field at the plasma boundary. This is a well-understood convex optimization problem with a closed-form solution. 

The optimization problem in \eqref{equation:cp:nescoil} is the basis of NESCOIL, which played important roles in the coil design of W7X and NCSX. However, the ill-posed nature of the problem means that the induction matrix $A_B$ is ill-conditioned. Without proper regularization, optima of \eqref{equation:cp:nescoil} typically contain unrealistically high peak current, as well as areas with dense, smoothly varying current that poorly approximate discrete coils.

Recent improvements in winding surface methods include better conditioning, regularization, sparsity promotion, linear physics objectives, and finite element bases. Boozer \cite{boozer_optimization_2000} performed truncated singular value decomposition (TSVD) on $A_B$ to remove small-scale, high-amplitude current modes with little impact on $f_B$. In the code REGCOIL, Landreman et al. \cite{landreman_improved_2017} introduces a Tikhonov regularization term,
\begin{equation}\label{equation:cp:regcoil}
    \begin{split}
        &\min_{\Phi'}[f_B(\Phi')+\lambda_{2}\int d^2a\|\K\|_2]
        = \min_{\Phi'}(\|A_B\Phi'+b_B\|_2 + \lambda_2\|A_K\Phi'+b_K\|_2), \\
        &\text{regularization weight }\lambda_2\geq0.
    \end{split}
\end{equation}
This modifies \eqref{equation:cp:nescoil} into a ridge regression problem, and effectively reduces the complexity of the resulting current. Elder \cite{elder_three-dimensional_2024} introduces L-1 regularization,
\begin{equation}\label{equation:cp:l1}
    \begin{split}
        &\min_{\Phi'}[f_B(\Phi')+\lambda_{1}\int d^2a\|\K\|_1] 
        = \min_{\Phi'}(\|A_B\Phi'+b_B\|_2 + \lambda_1\|A_K\Phi'+b_K\|_1), \\
        &\text{regularization weight }\lambda_1\geq0.
    \end{split}
\end{equation}
The new regularization term promotes sparsity in $\K$, and modifies \eqref{equation:cp:nescoil} into a lasso regression problem. Both regularization types are convex and can be applied to other linear functions of $\K$. Kaptanoglu et al.~\cite{kaptanoglu_topology_2024} introduce a non-convex, L-0 regularized current voxel method for coil topology optimization but its nonsmooth nature takes it outside the scope of this work. 

\begin{figure}
    \centering
    \subfloat[\centering]{
        \includegraphics[width=0.45\linewidth]{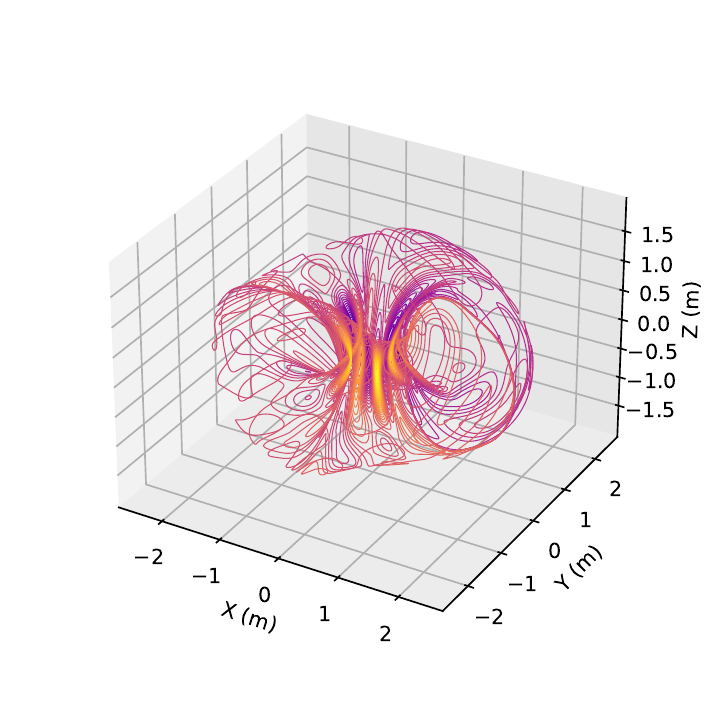} 
    %\caption{}
    }%
    \qquad
    \subfloat[\centering]{ 
        \includegraphics[width=0.45\linewidth]{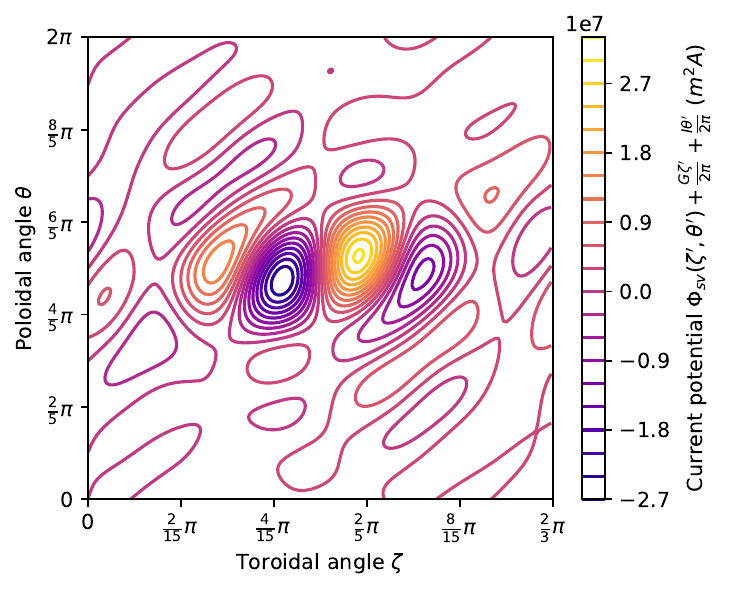} 
    %\caption{}
    }
    \caption{$\Phi'$ contours in $(X, Y, Z)$ (left) and $(\zeta, \theta)$ (right) of the NESCOIL optimum for NCSX, using an example winding surface $\approx2$ minor radii from the plasma. $\Phi'$ has the units of magnetic dipole moment density. Note the presence of both poloidal (links with the plasma) and windowpane (does not link with the plasma) contours.}
    \label{fig:cp:ncsx windowpane}
\end{figure}

These improvements highlight the two remaining shortcomings of existing winding surface methods. The first limitation is in the choice of objective. Both \eqref{equation:cp:regcoil} and \eqref{equation:cp:l1} are still limited to minimizing the L-2 or L-1 norm of linear functions of $\K$. The second limitation is that all existing winding surface methods are unconstrained. This makes it difficult to control the topology of the resulting surface current, beyond specifying the net poloidal current $G$ and net toroidal current $I$. Specifically, the NESCOIL and REGCOIL solutions usually contain a mix of poloidal, helical, and windowpane $\Phi'$ contours. This is especially likely on a compact device, as Fig. \ref{fig:cp:ncsx windowpane} shows. Because windowpane contours can be arbitrarily small, they can greatly complicate the coil-cutting process. To avoid unrealistic small windowpane coils, one often needs to trace and classify a large number of contours, or manually pick a set of realistic seeming coils. 

\subsection{QUADCOIL}
\label{section:theory:objectives}
Our algorithm, QUADCOIL, aims to address both drawbacks by allowing global optimization of a surface current by minimizing a non-convex quadratic objective under non-convex quadratic constraints. In the next section, we will discuss its formulation and characteristics.

QUADCOIL formulates the winding surface problem as a quadratically constrained quadratic program (QCQP):
\begin{equation}
\label{equation:problem:raw}
    \begin{split}
        &\min_{\Phi', a, b}\left[f(\Phi', a) + c^T b\right]\\
        &\text{subject to } \\
        & g_j(\Phi', a) + d_j^Tb \leq e_j, \\
        & h_k(\Phi', a) + m_k^Tb = n_k, \\
        &\text{for all } j = 1, ..., n_\text{ineq}, k = 1, ..., n_\text{eq}.
    \end{split}
\end{equation}
Here, $f$, $g_j$ are quadratic functions of $\Phi'$ and $a$. $n_\text{ineq}$ and $n_\text{eq}$ are the total number of inequality and equality constraints. $\Phi'\in\Rset^{n_{\Phi'}}$ is the discretized current potential, and $a\in\Rset^{n_a}, b\in\Rset^{n_b}$ are auxiliary variables for constructing complex problems. $c, d_j$ and $e_j$ are constants. Unlike in NESCOIL or REGCOIL, $f$ or $g_j$ may be convex or non-convex. This allows QUADCOIL to target a greater range of objectives and constraints.

By allowing non-convex quadratic objectives, QUADCOIL can target important engineering and physics objectives unavailable to existing winding surface methods. Examples include the stored magnetic energy in a coil set, coil-coil Lorentz force \cite{robin_minimization_2022}, the coil curvature proxy $\KK$, maximum dipole density, $\max_{\zeta', \theta'}|\Phi_\text{sv}|$, as well as coil-field alignment $\K\cdot \B$, potentially useful for estimating the critical current of high-temperature superconductor (HTS) coils. In this paper, we will focus on the quantity,
\begin{equation}\label{equation:KK}
    f^\infty_\kappa\equiv\max_{\zeta', \theta'}\|(\KK)_{(r, \Phi', z)}\|_\infty,
\end{equation}
as the primary example. This quantity is a quadratic proxy for the maximum curvature of the surface current, 
\begin{equation}
    \kappa_\text{max} = \max_{\zeta', \theta'} \frac{\|\KK\times\K\|_2}{\|\K\|_2^3},
\end{equation}
a non-linear function of $\Phi'$. Appendix \ref{appendix:KK} discusses its construction, and Appendix \ref{appendix:KK expr} contains its full expression in terms of $\Phi', \zeta'$ and $\theta'$.

With proper choice of $f, g_j, b$, and $a$, \eqref{equation:problem:raw} can represent a wide range of coil optimization problems. These include the REGCOIL problem \eqref{equation:cp:regcoil}, the L-1-regularized problem \eqref{equation:cp:l1}, and interesting new problems, such as minimizing the stored magnetic energy $E(\Phi')$ under a local error constraint:
\begin{equation}
\label{equation:problem:energy example}
    \begin{split}
        &\min_{\Phi'}E(\Phi')\\
        &\text{subject to } \\
        &|\B\cdot\hat \n| \leq B_{\text{max}}.
    \end{split}
\end{equation}

While a general QCQP is NP-hard, the exact global optimum of a``nearly convex'' QCQP can usually be found via its relaxation as a conic programming problem. This procedure is called Shor relaxation \cite{shor_quadratic_1987}, and is widely used for QCQPs outside stellarator optimization. Shor relaxation is uniquely suited for stellarator coil optimization because the primary consideration, field error, is often formulated as a convex quadratic term (i.e., $f_B$). This allows QUADCOIL to quickly find the global optimum of \eqref{equation:problem:raw} and avoid the time-consuming parameter tuning associated with constrained, non-convex optimization.

QUADCOIL combines Shor relaxation and a non-convex optimizer to perform rapid global optimization of stellarator coils. We discuss the properties of Shor relaxation, the implementation of QUADCOIL, the exactness of QUADCOIL solutions, as well as its limitations, in Section \ref{section:shor backgrounds}. The compatibility between Shor relaxation and the coil optimization problem will be further explored in \ref{section:numerical:parameter space}.

QUADCOIL applies to all linear discretizations of $\Phi'$. For simplicity, this paper uses the same Fourier representation as in REGCOIL and NESCOIL:
\begin{align}
\Phi'\left(\theta^{\prime}, \zeta^{\prime}\right)=\sum_j \Phi'_j\binom{\sin }{\cos }_j\left(m_j \theta^{\prime}-n_j \zeta^{\prime}\right),
\end{align}
where $j$ is a uniform index for the poloidal, toroidal mode numbers $(m, n)$, and the choice of $\sin$ and $\cos$. Similar to NESCOIL and REGCOIL, QUADCOIL supports all types of coil topology.

\subsection{Shor relaxation}
\label{section:shor backgrounds}
Shor relaxation rewrites a QCQP as a conic program (CP). This a well-studied class of convex optimization problems, and can be solved to arbitrary precision in polynomial time using interior point methods. A convex QCQP always has an exactly equivalent CP. While a non-convex QCQP does not have an exactly equivalent CP, it often has a CP that shares its global optimum or gives a good approximation to its global optimum. The detailed procedure for relaxing \eqref{equation:problem:raw} is summarized in Appendix \ref{appendix:shor procedures}.

There is a rich literature on the classification of QCQP with exact Shor relaxation, beyond the scope of this paper. Empirically, Shor relaxation typically works best when a problem is nearly convex. This is usually the case in winding surface optimization because the dominant factor in coil optimization is almost always the accuracy of the magnetic field, a convex quantity. Notably, besides $f_B$, many other forms of magnetic field objectives can be written as convex QCQPs. One example is the maximum local field error:
\begin{gather}
    f\text{ or }g = \max_{\zeta', \theta'} \|\B\cdot\n\|_2 \Leftrightarrow\\
    f\text{ or }g = a, \text{subject to }\\
    (\|\B\cdot\n\|_2)_{\zeta', \theta'} \leq a,\\
    a\geq0.
\end{gather}

While it is tricky to predict from $f, g_j, c_j$ and $d_j$ whether \eqref{equation:problem:raw} has an exact Shor relaxation, once the solution to the CP is obtained, it is easy to numerically test its exactness, requiring only the SVD of a $(n_{\Phi'}+n_a+1)\times (n_{\Phi'}+n_a+1)$ matrix. In practice, the CP of a typical QUADCOIL problem (40 $\Phi'$ harmonics, 1024 quadrature points per field period) takes only core-seconds to solve. Therefore, it is practical to solve the CP first and then test its exactness. We will discuss the mechanism of the optimality test in greater details in Appendix \ref{appendix:shor procedures}. In Section \ref{section:numerical:parameter space}, we will validate its effectiveness, and demonstrate the compatibility between Shor relaxation and stellarator coil optimization as part of a parameter space study.

Using these properties, QUADCOIL can substantially speed up constrained, non-convex winding surface problems. The algorithm works as follows:
\begin{enumerate}
    \item Upon receiving a QCQP coil optimization problem, QUADCOIL first solves its Shor relaxation and tests its exactness. 
    \item If the solution is exact, QUADCOIL returns the answer.
    \item If the solution is not exact, QUADCOIL initializes a local optimization solver, using the relaxation as the initial guess. 
\end{enumerate}
The implementation of QUADCOIL in this paper uses CVXPY \cite{agrawal2018cvxpy, diamond2016cvxpy}and MOSEK \cite{mosek} as the CP solver. Alternatively, when no constraint is present, \eqref{equation:problem:raw} can also be efficiently solved by local optimizers. The numerical example in Section \ref{section:numerical:KK} is one such unconstrained variation of \eqref{equation:problem:raw}. In Section  \ref{section:numerical:KK}, we will compare Shor relaxation to two popular unconstrained local optimization algorithms, ADAM \cite{kingma_adam:_2017}, a stochastic gradient descent algorithm, and L-BFGS-B \cite{byrd_limited_1995}, a quasi-Newton method. We will also briefly discuss the comparative advantage of each solver for different variations of \eqref{equation:problem:raw}.

For simplicity, in this paper, we have only implemented step 3 for unconstrained problems using L-BFGS-B. The choice and tuning of a constrained local optimization solver for this step requires further work. However, Section\ref{section:numerical:parameter space} shows that even without step 3, approximate solutions from QUADCOIL can still predict filament coil complexity.

\begin{figure}
    \centering
    \subfloat[\centering]{
        \includegraphics[width=0.45\linewidth]{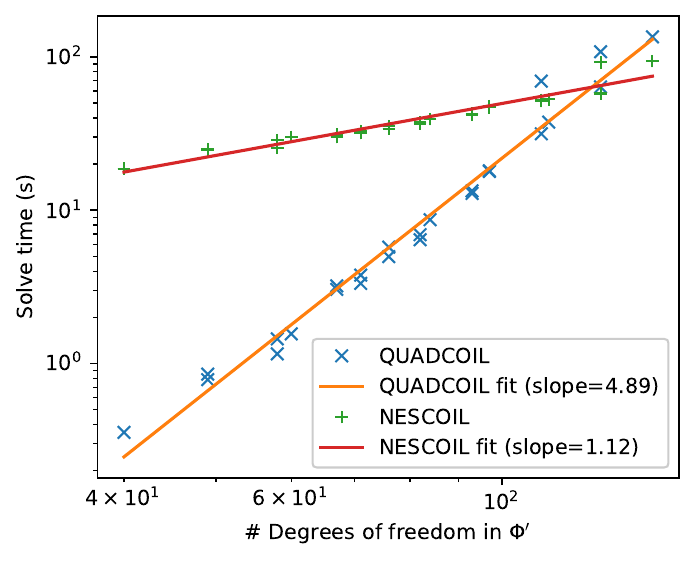} 
    %\caption{}
    }%
    \qquad
    \subfloat[\centering]{ 
        \includegraphics[width=0.45\linewidth]{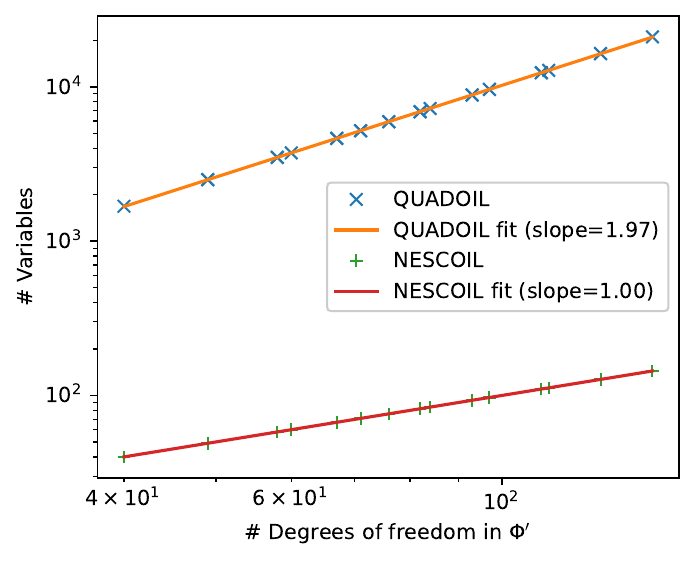} 
    %\caption{}
    }
    \caption{Time $t$ required to solve \eqref{equation:num:no windowpane} (left) and the number of free variables (right) for NCSX as a function of $n_{\Phi'}$. Note the power-of-two difference in the numbers of free variables in QUADCOIL and NESCOIL.}
\label{fig:theory:runtime}
\end{figure}

Shor relaxation comes with one drawback. The number of free variables in the relaxed CP scales as $\mathcal{O}[(n_{\Phi'}+n_a)^2 + n_{b}]$. As a result, compared to existing current potential methods, QUADCOIL scales poorly with high $n_{\Phi'}$ and $n_a$. This may become a problem in future works applying finite element basis to QUADCOIL. See Fig \ref{fig:theory:runtime} for the solve time of \eqref{equation:num:KK original} in QUADCOIL as a function of $n$. ($m=0$ in this problem). 
\subsection{Generating winding surfaces}
Existing winding surface codes typically generate the winding surface by uniformly offsetting the plasma surface along the normal direction. For a highly shaped plasma surface, this method can produce a winding surface with self-intersections, regions with very large principal curvatures, and unevenly spaced quadrature points. This is especially detrimental to the minimization of maximum curvature, and potentially, Lorentz force.

\begin{figure}
    \centering
    \subfloat[\centering]{
        \includegraphics[width=0.45\textwidth]{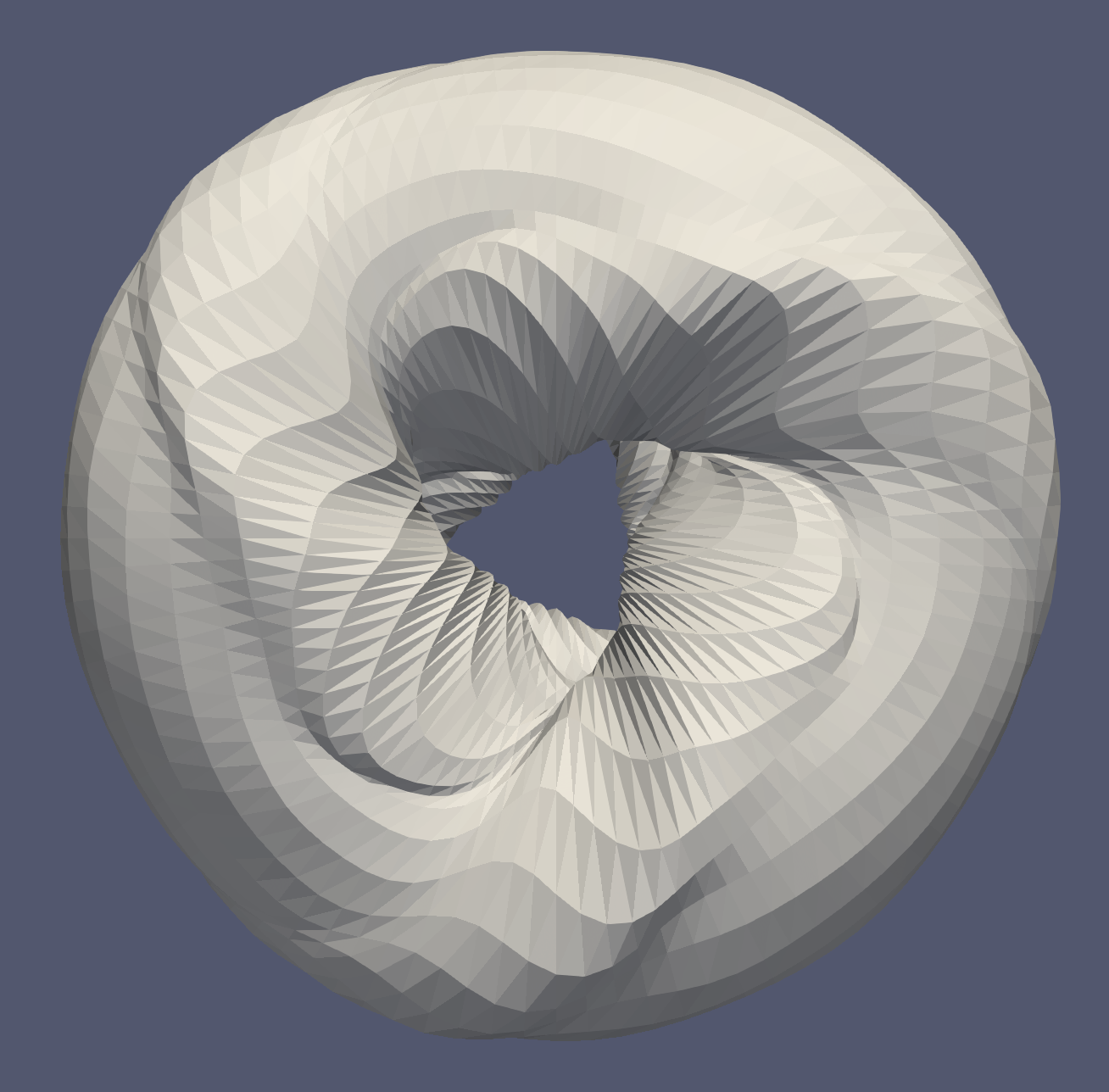} 
    %\caption{}
    }%
    \qquad
    \subfloat[\centering]{ 
        \includegraphics[width=0.45\textwidth]{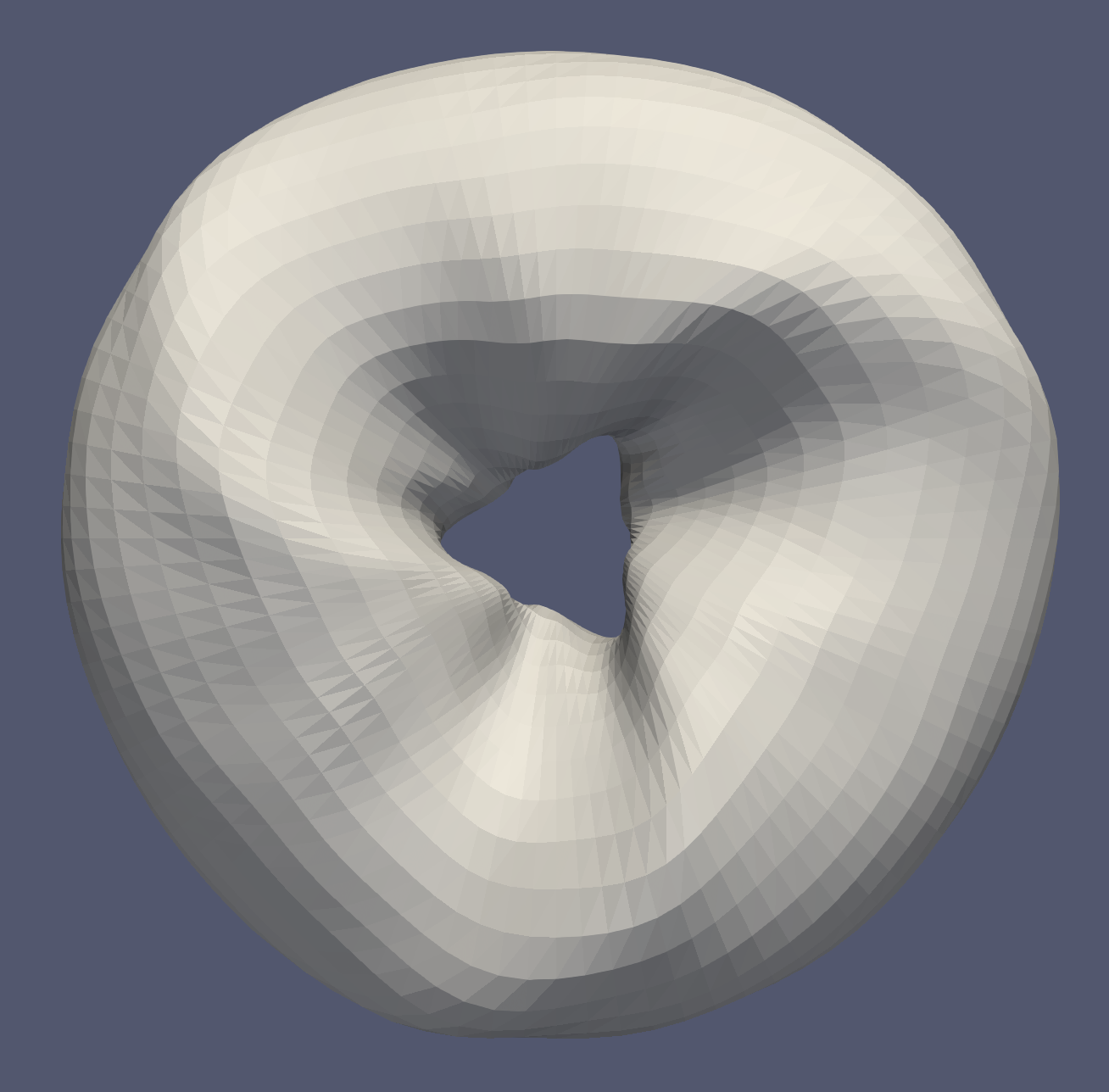} 
    %\caption{}
    }
    \caption{NCSX winding surface generated by uniformly offsetting quadrature points on the plasma surface along the normal direction (left) vs. surface generated by QUADCOIL (right).}
    \label{fig:appendix:extend}
\end{figure}

QUADCOIL generates a winding surface with the following procedure:
\begin{enumerate}
    \item Offset quadrature points on the plasma surface along the unit normal by $d_{\text{cs}}$. 
    \item Evaluate the cross sections of the surface at uniform intervals in the Cartesian toroidal angle, $\zeta'$.
    \item Find the convex hull of the cross-section.
    \item Fit a periodic cubic spline to the convex hull and redefine quadrature points at equal arc lengths.
    \item Fit a surface to the quadrature points.
\end{enumerate}
The poloidal coordinate on the winding surface, $\theta'$, is proportional to the arc length of the cubic spline. We choose the zeros of $\theta'$ as the outboard point with $Z$ closest to the splines' centers of weight. Fig.\ref{fig:appendix:extend} compares our method with the more common uniform offset method. Note that our method creates a smoother surface with no self-intersections and uniformly spaced quadrature points. 

This method for generating winding surfaces has one potential drawback. Empirically, the field error in a stellarator is the most sensitive to the coil-plasma distance on the concave regions of the plasma surface. Using the convex hull of cross sections, therefore, can potentially increase the optimal field error. Yet, one may argue that a physical vessel with no concave region is simpler to engineer. The method for producing the optimal winding surface is beyond the scope of this paper. As Section \ref{section:numerical:parameter space} shows, our method is sufficiently well-behaved across a variety of equilibria and optimization targets for QUADCOIL to serve as a coil curvature proxy. 

% \section{Convex hull of the non-convex constraint}
% \label{appendix:hull}

\section{Numerical examples}
\label{section:numerical}
In this section, we demonstrate QUADCOIL's potential as an initial state tool and coil complexity proxy with three numerical examples. In \ref{section:numerical:topology}, we demonstrate an inequality constraint that gives the option to eliminate all windowpane $\Phi'$ contours and substantially simplify the coil-cutting process. In \ref{section:numerical:KK}, we show that QUADCOIL can outperform REGCOIL in coil simplification by directly targeting a non-convex penalty. In \ref{section:numerical:parameter space}, we validate QUADCOIL's effectiveness as a coil complexity proxy. We observe a correlation between the optimal normalized field error produced by QUADCOIL and filament optimizations across 4436 combinations of different equilibrium, maximum curvature, coil-coil spacing, and coil-plasma spacing. \cite{lanke_quadcoil_data}

\subsection{Topology control}
\label{section:numerical:topology}
QUADCOIL gives the option to create a current configuration with purely poloidal $\Phi'$ contours. This is achieved by setting $G=0$, and solving: 
\begin{gather}
    \label{equation:num:no windowpane}
    \min_{\Phi'} f_B(\Phi'),\\ \label{equation:K_no_windowpane}
    \text{subject to }
    K_\theta'(\Phi') \text{sign}(I)\geq0.
\end{gather}
Here, $\text{sign}(I)$ is the sign of the net poloidal current. \eqref{equation:K_no_windowpane} is a necessary condition for all $\Phi'$ contours to be poloidal. It prevents the sheet current $\K$ from flowing against the direction of the net poloidal current, effectively forbidding windowpane currents. 

A similar constraint can be constructed for helical coils:
\begin{align}
\K'(\Phi')\cdot \hat I_{\text{helical}} \geq0,
\end{align}
where $\hat I_{\text{helical}}$ is a unit vector tangent to the direction of the net helical current.

It is difficult to obtain a similar collection of poloidal current paths with REGCOIL because the REGCOIL penalty does not directly target the sign of poloidal current. The first method is to evaluate a large number of contours during the coil-cutting process and choose a subset that has the desired topology and provides a sufficiently accurate magnetic field. This is computationally costly and often requires human input. Alternatively, one can search for a regularization parameter $\lambda_{2}$ in \eqref{equation:cp:regcoil} that produces the lowest-$f_B$ current configuration with the desired topology, although there is no guarantee a given topology will be achieved. Moreover, this requires a large number of REGCOIL solves and can converge slowly without a good initial guess for $\lambda_{2}$. 

\begin{figure}
    \centering
    \subfloat[\centering][
        QUADCOIL
    ]
    {
        \label{fig:num:no windowpane QUADCOIL}
        \includegraphics[width=0.4\textwidth]{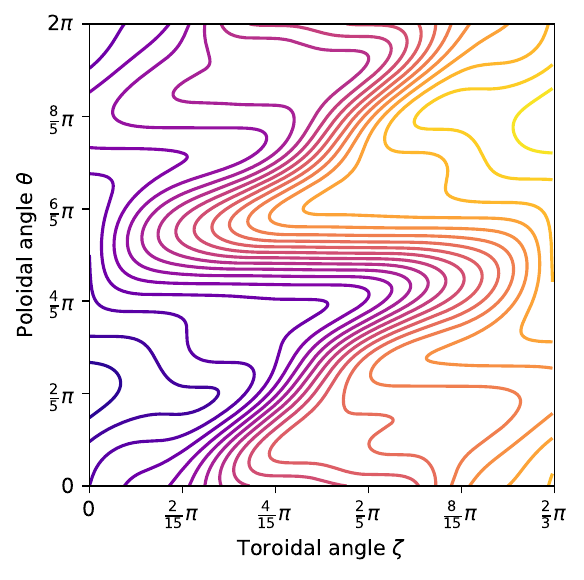} 
    }%
    \qquad
    \subfloat[\centering][
        Binary search with REGCOIL
    ]{  
        \label{fig:num:no windowpane REGCOIL}
        \includegraphics[width=0.5\textwidth]{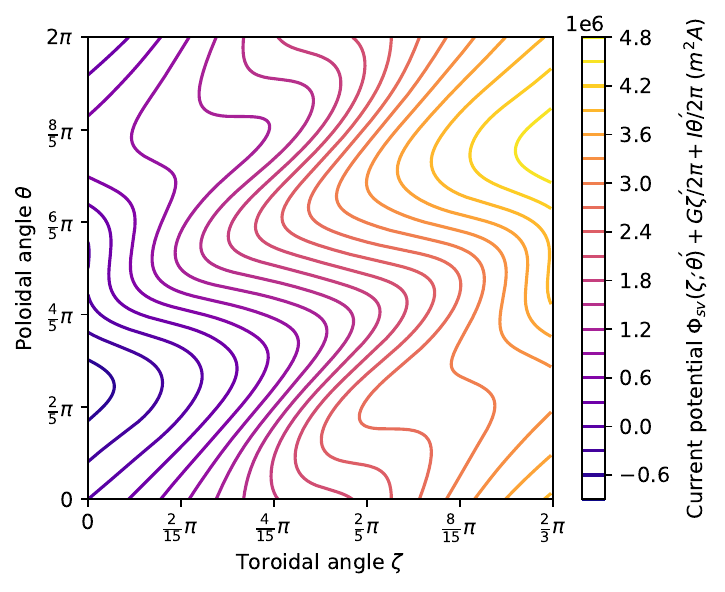} 
    }
    \caption{$\Phi'$ contours in $(\zeta, \theta)$ of the lowest-$f_B$ QUADCOIL and REGCOIL configurations with no windowpane current. The REGCOIL configuration has higher field error, because the Tikhnov regularization does not directly target the sign of the poloidal current, and therefore overconstrains.}
    \label{fig:num:no windowpane}
\end{figure}

Compared to REGCOIL, QUADCOIL can produce a purely poloidal current configuration with lower $f_B$ in a fraction of the computation time. Fig. \ref{fig:num:no windowpane} compares two such current configurations for NCSX\cite{nelson_design_2003}, obtained with QUADCOIL and REGCOIL. The winding surface is located 2 minor radii away from the plasma. We obtain the configuration in Fig. \ref{fig:num:no windowpane QUADCOIL} by solving \eqref{equation:num:no windowpane} in QUADCOIL, without any additional regularization. This configuration has $f_B=0.607$ T$^2$m$^2$, and requires $1.77$s to solve. We obtain the configuration in Fig. \ref{fig:num:no windowpane REGCOIL} by performing a binary search in $\log_{10}(\lambda_{2})$, to find the lowest-$f_B$ REGCOIL solution satisfying $K_\theta'(\Phi') \text{sign}(I)\geq0.$. The search starts at $\log_{10}(\lambda_{2}) = -50 \text{ and } 1$, and terminates when the relative change in $f_B$ falls below $0.1\%$. The binary search requires 15 REGCOIL solves and $43.50$ s to complete, substantially slower than QUADCOIL. The configuration has $f_B=0.792$T$^2$m$^2$, $30.4\%$ higher than the QUADCOIL value. 

\subsection{Nonconvex penalty}
\label{section:numerical:KK}
In this section, we show that QUADCOIL can outperform REGCOIL when directly targeting a non-convex objective. In this section, we design helical coils for HSX, and modular coils for NCSX and W7X, by solving
\begin{equation}
\label{equation:num:KK original}
        \min_{\Phi'}\left[
            f_B(\Phi')
            + \lambda_\kappa f^\infty_\kappa(\Phi')
        \right].
\end{equation}
In all 3 cases, the winding surfaces are located 2 minor radii from the plasma.

\begin{figure}
    \centering
    \includegraphics[width=1\linewidth]{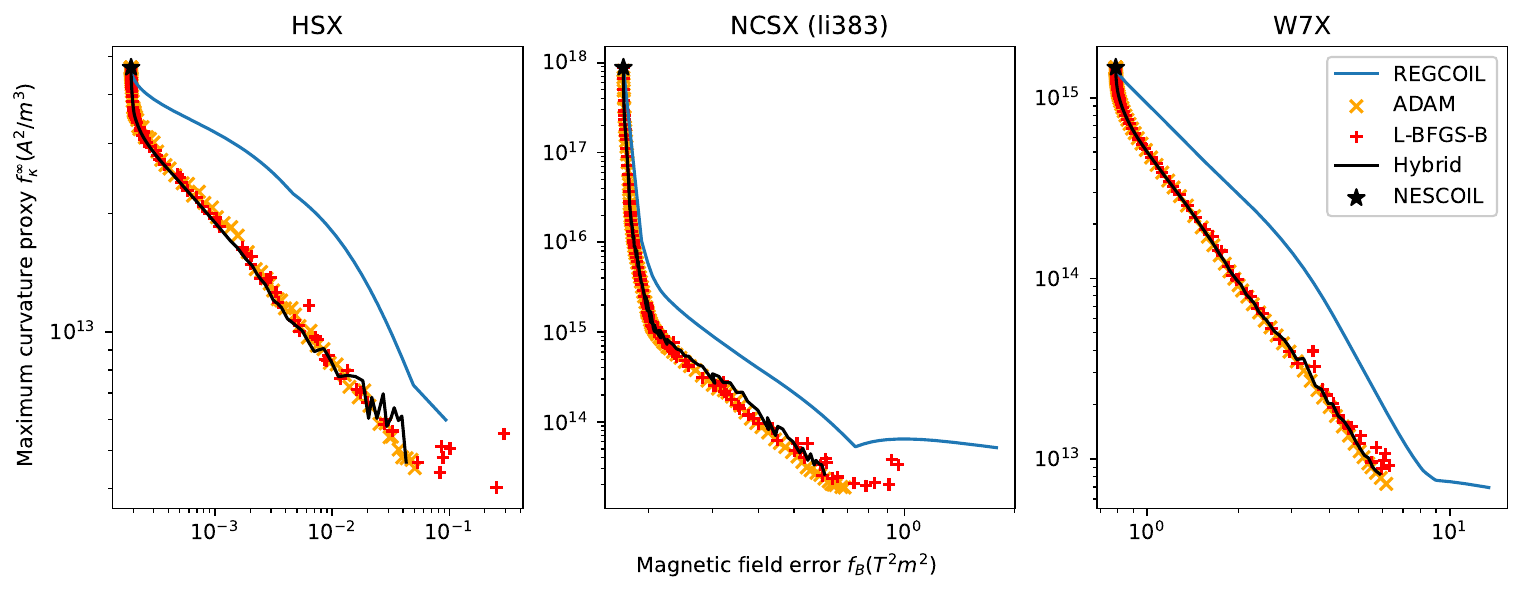}
    \caption{Trade-off plots between the squared flux $f_B$ and the curvature proxy $f_\kappa^\infty$. This figure includes 150 REGCOIL runs, and 100 runs each for QUADCOIL using the hybrid, ADAM, and L-BFGS-B solvers. The REGCOIL runs sweeps $\lambda_2$ exponentially. The QUADCOIL runs solve \eqref{equation:num:KK original}, sweeping $\lambda_\kappa$ exponentially. For ADAM or L-BFGS-B, each marker represents a scan point. For visual clarity, the REGCOIL and QUADCOIL hybrid scan points are not marked.}
    \label{fig:num:KK}
\end{figure}

QUADCOIL consistently outperforms REGCOIL when targeting specific non-convex penalties. Fig. \ref{fig:num:KK} compares the trade-off between $f_B$ and $f_\kappa^\infty$ across REGCOIL and QUADCOIL using 3 different solvers:
\begin{enumerate}
    \item A hybrid solver combining Shor relaxation and L-BFGS-B, as discussed in \ref{section:shor backgrounds}. 
    \item Directly solving with L-BFGS-B, the quasi-Newton algorithm.
    \item Directly solving with ADAM, the stochastic gradient descent algorithm.
\end{enumerate}
As Fig. \ref{fig:num:KK} shows, with all 3 solvers, QUADCOIL consistently produces lower $f^\infty_\kappa$ than REGCOIL at equal $f_B$. Note that while a single QUADCOIL case runs in a similar time as a single REGCOIL case, producing a configuration with a target $f_B$ requires a costly $\lambda_{2}$ search similar to the example in Section \ref{section:numerical:topology}. Therefore, the actual time required to obtain a desired current configuration using REGCOIL can be significantly longer than the per-case solve time reported in Table~\ref{tab:num:KK time}. 

We now compare the 3 different solvers for unconstrained QUADCOIL. While all 3 solvers yield comparably accurate results, ADAM is $5-20\times$ slower than the hybrid solver and L-BFGS-B. This suggests that the landscape of \eqref{equation:num:KK original} may be sufficiently smooth and nearly convex, that stochastic gradient descent is not necessary. As discussed in Section\ref{section:shor backgrounds}, directly solving with L-BFGS-B will also likely scale better than Shor relaxation to a higher number of degrees of freedom. Therefore, we believe that the L-BFGS-B solver is preferable at high resolutions, or in regions of the parameter space where the non-convexity is known to be strong. Conversely, the hybrid solver is preferable at low resolutions, or when the objective is known to be convex or sufficiently near convex. This comparison may change with an improved implementation of the hybrid solver.

In the next section, we will demonstrate the effectiveness of $f_\kappa^\infty$ as a proxy for the maximum curvature in filament optimization, by solving a constrained optimization problem.

\begin{table}
    \centering
    \begin{tabular}{|c|c|c|c|c|} \hline 
         &REGCOIL & QUADCOIL (hybrid)&   QUADCOIL (L-BFGS-B)&QUADCOIL (ADAM)\\ \hline 
         HSX&  \makecell{$12.6 - 12.7$ s, \\ Med$(t) = 12.7$ s} &  \makecell{$0.4 - 15.8$ s,\\ Med$(t) = 1.4$ s} &  \makecell{$1.0 - 19.3$ s, \\ Med$(t) = 3.4$ s} & \makecell{$5.0 - 32.7$ s, \\ Med$(t) = 10.5$ s}\\ \hline 
 NCSX& \makecell{$9.5 - 10.9$ s, \\ Med$(t) = 9.5$ s  } &  \makecell{$0.2 - 13.1$ s,\\ Med$(t) = 4.7$ s} &  \makecell{$2.3 - 12.5$ s, \\ Med$(t) = 5.0$ s} & \makecell{$9.6 - 73.4$ s, \\ Med$(t) = 25.8$ s}\\ \hline 
         W7X&  \makecell{$16.0 - 18.7$ s, \\ Med$(t) = 16.0$ s} &  \makecell{$0.3 - 12.5$ s,\\ Med$(t) = 1.5$ s} &  \makecell{$1.2 - 16.9$ s, \\ Med$(t) = 3.6$ s} & \makecell{$9.8 - 63.2$ s, \\ Med$(t) = 34.9$ s}\\ \hline
    \end{tabular}
    \caption{Minimum, maximum, and median solve time comparison of QUADCOIL, L-BFGS-B, and ADAM minimizing \eqref{equation:num:KK original}. Note that the REGCOIL solve times in this table are per case. To find a REGCOIL solution with a target $f_B$ will require a $\lambda_{2}$ search, and will be substantially slower than the numbers in this table.}
    \label{tab:num:KK time}
\end{table}

\subsection{Validating QUADCOIL as a complexity metric}
\label{section:numerical:parameter space}

In this section, we present a parameter space study showing that QUADCOIL can predict the optimum normalized field error,
\begin{equation}
    \label{eq:JB}
    J_B \equiv \frac{1}{2} \frac{\int_{\text{plasma}}\left(\mathbf{B}_\text{coil} \cdot \n-B_T\right)^2 d^2s}{\int_{\text{plasma}}|\mathbf{B}|^2 d^2s},
\end{equation}
in a filament optimization for a given combination of maximum curvature $\kappa_0$, minimum coil-coil spacing $d_{\text{cc}}^{\min}$, and minimum coil-plasma spacing $d_{\text{cs}}^{\min}$. 
\subsubsection{Filament models}
Before constructing a QUADCOIL proxy, we first introduce the two filament models used in our parameter space study. In both models, all coils have equal, fixed current $I_\text{coil}$ based on the equilibrium poloidal flux. The first model constrains the coils to a prescribed winding surface. We refer to this model as the CWS (coil on winding surfaces) model. The second model allows the filament to move freely in space, as long as the coil-plasma spacing is greater than the specified threshold $d_{\text{cs}}^\text{min}$. We refer to this model as the free filament model. While the free model is more realistic, the CWS model's geometry bears more similarity to the sheet current $\textbf{K}$ in the winding surface approach. The objective functions are, respectively:
\begin{gather}
    J_\text{CWS} \equiv 1J_B+ 1000 J_{\kappa}  + 1 J_{l} + 500 J_{\text{cc}},\label{equation:filament objectives cws}\\
    J_\text{free} \equiv 10J_B+ 500 J_{\kappa}  + 1 J_{l} + 100 J_{\text{cc}} + 100J_{\text{cs}}.\label{equation:filament objectives free}
\end{gather}
The 5 penalty terms are:
\begin{enumerate}
    \item $J_B$, the normalized squared flux, as defined in \eqref{eq:JB}.
    \item $J_{\kappa}$, the maximum curvature constraint:
    \begin{equation}
            J_{\kappa} = a^2\sum_{\text{coils}}\frac{1}{2} \int_{\text{curve}} \text{max}(\kappa - \kappa_0, 0)^2 ~dl.
    \end{equation}
    Here, $\kappa_0$ is the target maximum curvature.
    \item $J_l$, the maximum coil length constraint:
    \begin{equation}
        J_l = \frac{1}{a_{\text{WS}}^2}\sum_{i = 1}^{N_\text{coil}}0.5\max(l_i - l_{\text{target}}, 0)^2.
    \end{equation}
    Here, $l_{\text{target}}$ is a desired threshold for the maximum coil length, and $a_{\text{WS}}$ is the effective minor radius of the winding surface. 
    \item  $J_{\text{cc}}$, the coil-coil spacing constraint:
    \begin{gather}
            J \equiv \frac{1}{d^\text{ref}_{\text{cc}}}\sum_{i = 1}^{N_\text{coil}} \sum_{j = 1}^{i-1} \int_{\text{curve}_i} \int_{\text{curve}_j} \max(0, d^{\min}_{\text{cc}} - \| \mathbf{r}_i - \mathbf{r}_j \|_2)^2 ~dl_j ~dl_i,\\
            d^\text{ref}_{\text{cc}} \equiv \frac{2\pi(R_\text{WS}-a_\text{WS})}{N_\text{coil}}.
    \end{gather}
    $R_\text{WS}$ is the effective major radius of the winding surface. Intuitively, the reference distance $d^\text{ref}_{\text{cc}}$ is the effective inboard circumference of the winding surface, divided by the number of coils.
    \item $J_{\text{cs}}$, the coil-plasma spacing constraint:
    \begin{equation}
        J_{\text{cs}} \equiv \frac{1}{a} \sum_{i = 1}^{N_\text{coil}} \int_{\text{curve}_i} \int_{\text{surface}} \max(0, d^{\min}_{\text{cs}} - \| \mathbf{r}_i - \mathbf{s} \|_2)^2 ~dl_i ~ds.
    \end{equation}
\end{enumerate}
In both filament models, we choose the coil number per half-field-period closest to satisfying
\begin{equation}
    \frac{N_{\text{coil}}}{\text{aspect ratio}}\approx 4.65.
\end{equation}
To keep the coil-coil spacing consistent across different equilibria. The target, $4.65$, is the value of this ratio measured in W7X. For faster convergence, all constraints in our filament optimization are enforced by penalty functions. As a result, they can be broken when multiple conflicting constraints are present, or the gradient of $J_B$ is large.

Now, we construct the equivalent QUADCOIL problem to \eqref{equation:filament objectives cws} and \eqref{equation:filament objectives free}:
\begin{equation}
    \label{equation:problem:equivalent}
    \begin{split}
        \min_{\Phi'}&f_B, \\
        \text{subj}&\text{ect to}\\
        \max_{(\zeta', \theta')} |K|^2 \leq K_\text{max}^2,\quad
        \frac{f_\kappa^\infty}{K_\text{max}^2} &\leq \kappa_0,\quad
        K_\text{max}\equiv \frac{I_{\text{coil}}}{d^{\text{min}}_{\text{cc}}}.
    \end{split}   
\end{equation}
Here, the first constraint emulates the coil-coil spacing by limiting the maximum permitted current density. The second constraint approximates the maximum curvature limit by dividing $f_\kappa^\infty$ by the square of the target maximum current. This has a similar functional form of the curvature of a filament coil:
\begin{align}
\kappa = \frac{\mathbf{I}\cdot\nabla\mathbf{I}}{|\mathbf{I}|^2}.
\end{align}
\begin{figure}
    \centering
    \includegraphics[width=0.8\linewidth]{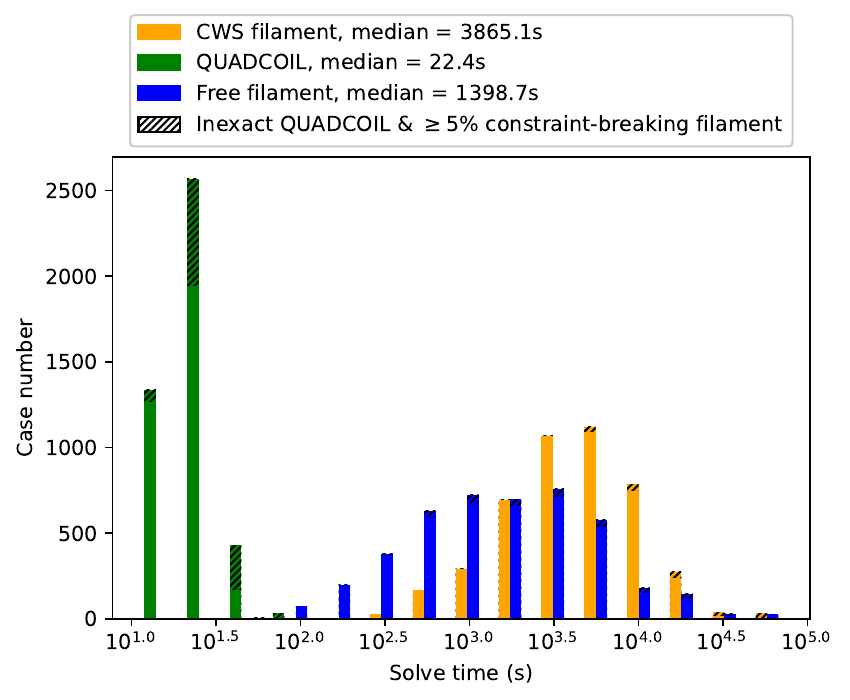}
    \caption{Run time comparison of QUADCOIL, CWS filament and free filament. The median time of QUADCOIL is $\approx10^2\times$ faster than both CWS filaments and free filaments. Note that the constraints in filament optimization are enforced by penalty functions, and can be broken when multiple conflicting constraints are present, the gradient of $J_B$ is large, or the winding surface has large principal curvatures.}
    \label{fig:num:param:time}
\end{figure}
\subsubsection{Numerical setup}
Our parameter scan uses $38$ stellarator and heliotron equilibria from the equilibrium dataset in \cite{kappel_magnetic_2024, kappel_magnetic_data}. For each equilibrium, we optimize the normalized field error $J_B$ for 125 combinations of $\kappa_0$, $d_{\text{cc}}^{\min}$, and $d_{\text{cs}}^{\min}$ targets. 

The CWS filament optimization uses the accessibility branch \cite{simsopt_accessibility} of Simsopt \cite{landreman_simsopt:_2021}. We represent the $(\zeta', \theta')$ coordinate of a filament constrained to a surface with a Fourier series:
\begin{equation}
\begin{alignedat}{2}
    \zeta'(t) &=  &&\sum_{m=0}^{\text{order}} \zeta'_{c,m}\cos(m t) + \sum_{m=1}^{\text{order}} \zeta'_{s,m}\sin(m t), \\
            \theta'(t) &= 2\pi t + &&\sum_{m=0}^{\text{order}} \theta'_{c,m}\cos(m t) + \sum_{m=1}^{\text{order}} \theta'_{s,m}\sin(m t) .\\
\end{alignedat}
\end{equation}
We first perform the optimization with $\text{order}=2$. When it converges, we incrementally increase the order to $4$, $6$ and $8$, and repeat the optimization using the low-order optimum as the initial condition. This is a standard practice in filament optimization. Filament optimization is non-convex. Therefore, it is very sensitive to the initial state and can be trapped at a local minimum. Directly initializing a high-order filament optimization using circular coils can lead to poor convergence. Initializing a low-order filament optimization with circular coils, and then gradually increasing the order of the curve often improves the solve time and the optimal value of the objective function. 

We represent a free filament with a Fourier series:
\begin{equation}
    \begin{split}x(t) &= \sum_{m=0}^{\text{order}} x_{c,m}\cos(mt) + \sum_{m=1}^{\text{order}} x_{s,m}\sin(mt) \\
y(t) &= \sum_{m=0}^{\text{order}} y_{c,m}\cos(mt) + \sum_{m=1}^{\text{order}} y_{s,m}\sin(mt) \\
z(t) &= \sum_{m=0}^{\text{order}} z_{c,m}\cos(mt) + \sum_{m=1}^{\text{order}} z_{s,m}\sin(mt).
\end{split}
\end{equation}
Similar to the constrained filaments, we first perform the optimization with $\text{order}=4$. Upon convergence, we repeat the optimization at $\text{order}=6$, and then $8$, similar to the CWS filaments. To make sure all filament cases have approximately equal coil spacing, we choose the coil number per half period closest to satisfying:
$$
\frac{\text{total coil \#}}{\text{aspect ratio}} \approx 4.65 \text{ (the W7X value)}.
$$

Filament methods are sensitive to their initial states. For consistency, we initialize the constrained filaments as uniformly-spaced, cross sections of the winding surface ($\zeta'_{c,m}, \zeta'_{s,m}, \theta'_{c,m}, \theta'_{s,m}=0$), and free filaments by fitting $\text{order}=4$ space curves to these cross sections using inverse Fourier transform.

For consistency across all equilibria, we normalize $\kappa_0$, $d_{\text{cc}}^{\min}$, $d_{\text{cs}}^{\min}$, and $l^{\max}$ into 4 dimensionless parameters using their associated scale lengths:
\begin{enumerate}
    \item $k \equiv \frac{\kappa_0^{-1}}{a}$ is the ratio between the radius of curvature and the minor radius. 
    \item $d_{\text{cc}}^{\text{norm}} \equiv \frac{d_{\text{cc}}^{\min}}{d^\text{ref}_{\text{cc}}}$ is the coil-coil spacing, normalized by the reference distance $d^\text{ref}_{\text{cc}}$. 
    \item $d_{\text{cs}}^{\text{norm}} \equiv \frac{d_{\text{cs}}^{\min}}{a}$ is the coil-plasma spacing normalized by the minor radius.
    \item $l^{\text{norm}} \equiv \frac{l^{\max}}{2\pi a_{WS}}$ is the total coil length normalized by an effective poloidal circumference, $2\pi a_{WS}$.
\end{enumerate}

We perform grid scans in $(k, d_{\text{cc}}^{\text{norm}}, d_{\text{cs}}^{
\text{norm}})$. The choice of scan points is based on the measured values from NCSX and W7X. Empirically, a coil length limit $l^{\text{norm}}$ is often necessary for free filament optimization to converge. Since QUADCOIL cannot target this quantity, we choose the constant value of $l^{\text{norm}}=1.5$ in all cases. Both filament models use the L-BFGS-B algorithm. 
\begin{table}
    \centering
    \begin{tabular}{|c|c|c|} \hline 
         Scan points&  NCSX values& W7X values\\ \hline 
         $k \in \{0.1,0.2,0.3,0.4,0.5\}$&  $0.30$& $0.33$\\ \hline 
         $d_{\text{cc}}^{\text{norm}}  \in \{0.2,0.4,0.6, 0.8, 1.0\}$&  $0.65$& $0.51$\\ \hline 
         $d_{\text{cs}}^{\text{norm}} \in \{1.0,1.5,2.0,2.5,3.0\}$&  $1.73$& $2.45$\\ \hline 
         $l^{\text{norm}}=1.5$&  $1.43$& $1.13$\\ \hline
    \end{tabular}
    \caption{Values of $(k, d_{\text{cc}}^{\text{norm}}, d_{\text{cs}}^{\text{norm}}, l^{\text{norm}})$ in the grid scan, compared with measured values at NCSX and W7X. $R_{\text{WS}}, a_{\text{WS}}$ values in this table are estimated by finding best-fit surfaces that interpolate all NCSX coils and non-planar W7X coils.}
    \label{tab:my_label}
\end{table}

In QUADCOIL, we use 4 $\sin$ and $4$ $\cos$ modes each in both the $\zeta'$ and $\theta'$ directions for $\Phi'(\zeta', \theta')$. For simplicity of implementation, in Section \ref{section:numerical:parameter space}, we will solely rely on the Shor relaxation part of QUADCOIL. As we will show in the next section, QUADCOIL is often effective even when the exactness test \eqref{equation:appendix:test} fails. As a control group, we will also compare QUADCOIL and NESCOIL results using the same winding surfaces.

\subsubsection{Results}
\label{section:numerical:parameter space:results}
\begin{figure}
    \centering
    \subfloat[\centering]{
        \includegraphics[width=7cm]{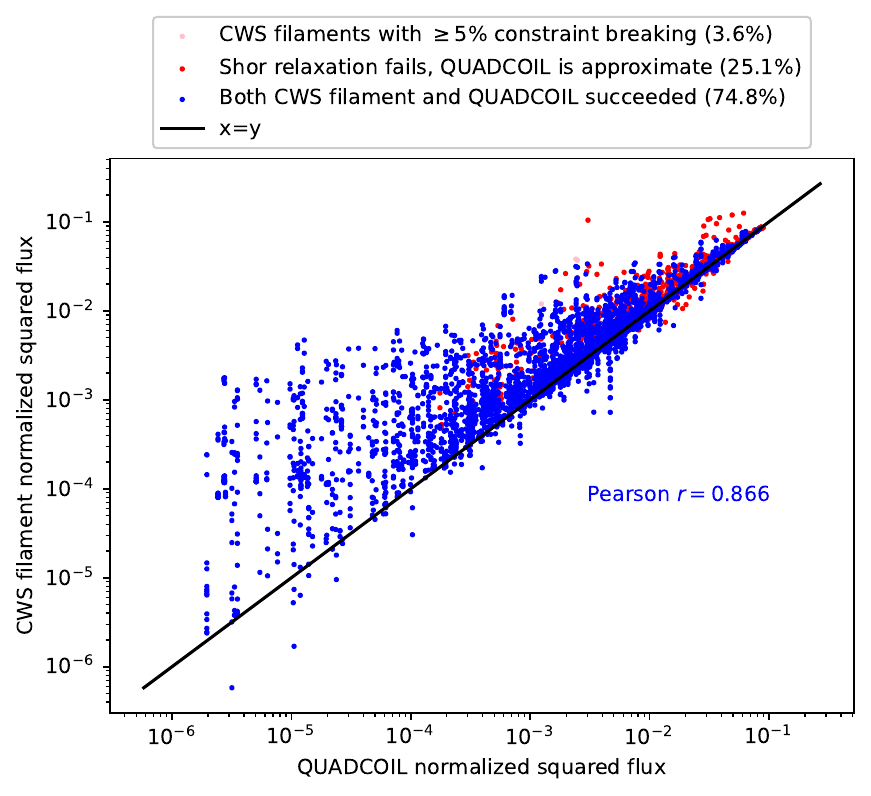} 
    %\caption{}
    }%
    \qquad
    \subfloat[\centering]{ 
        \includegraphics[width=7cm]{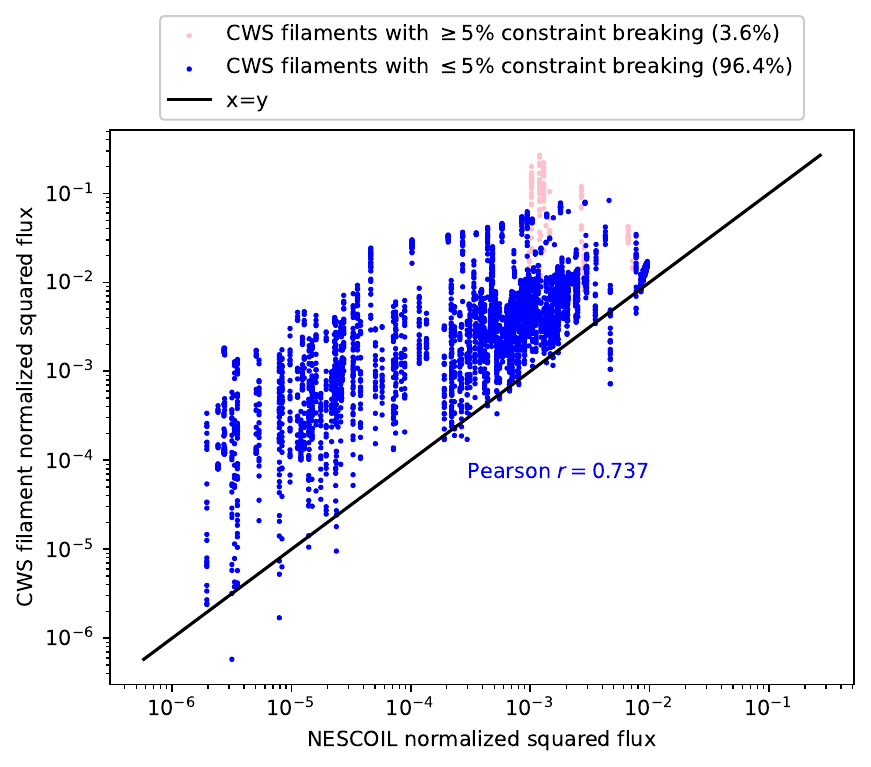} 
    %\caption{}
    }
    \caption{QUADCOIL $J_B$ v. CWS filament $J_B$ (left) and NESCOIL $J_B$ v. CWS filament $J_B$ (right). The correlation is stronger in the left plot, indicating QUADCOIL is a more effective coil complexity proxy than NESCOIL.}
    \label{fig:num:param:cws}
\end{figure}

Across all converged combinations of equilibria, $\kappa_0$, $d_{\text{cc}}^\text{min}$ and $d_{\text{cs}}^\text{min}$, we find the $J_B$ values measured from both CWS and free filament optimization correlate with the values from QUADCOIL optima. Fig. \ref{fig:num:param:cws} shows that CWS filament coils rarely achieve lower $J_B$ than QUADCOIL. The correlation between $J_B$ from the CWS filament method and NESCOIL is weaker in comparison. This suggests that QUADCOIL may serve as an empirical lower bound of the best normalized field error achievable by CWS filament coils. The spread of points in Fig.\ref{fig:num:param:cws} suggests that the correlation is stronger at higher values of $J_B$. This indicates that QUADCOIL is best suited for excluding equilibria that are hard to build coils for. As Fig.\ref{fig:num:param:time} shows, QUADCOIL cases have a median run-time of $22.4$s, $175\times$ faster than the median run time of $3865.1$s of CWS filaments.

\begin{figure}
    \centering
    \subfloat[\centering]{
        \includegraphics[width=7cm]{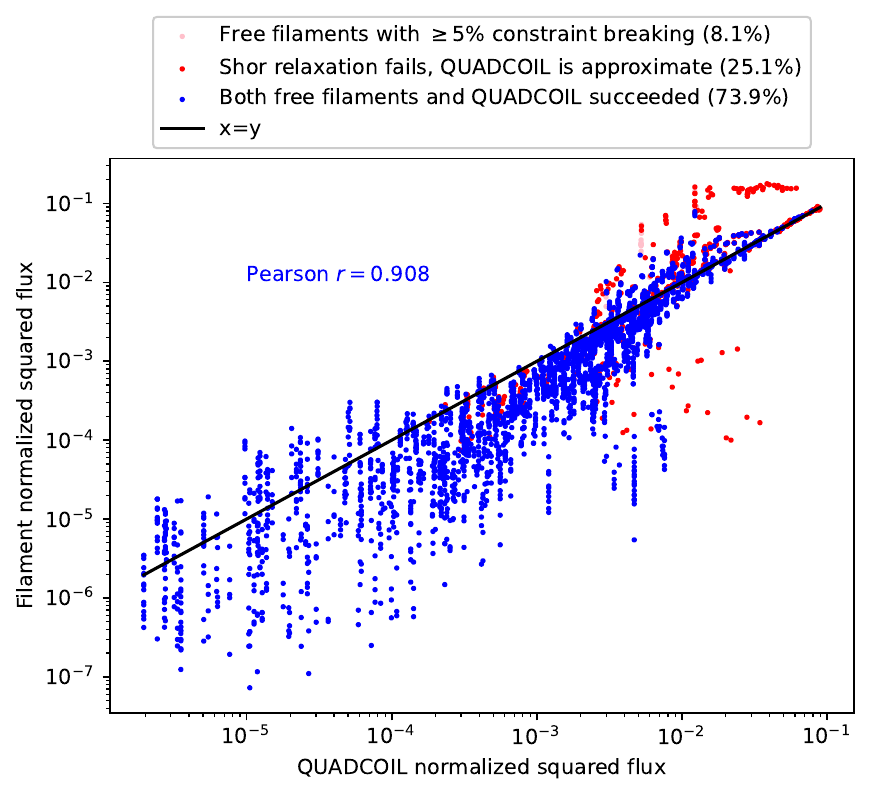} 
    %\caption{}
    }%
    \qquad
    \subfloat[\centering]{ 
        \includegraphics[width=7cm]{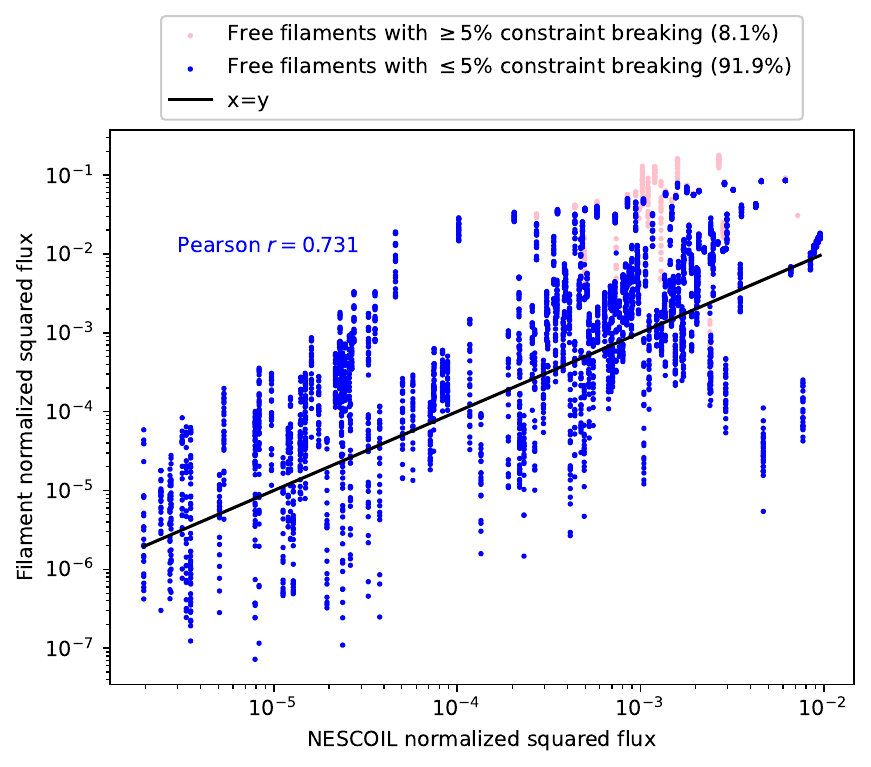} 
    %\caption{}
    }
    \caption{QUADCOIL $J_B$ v. free filament $J_B$ (left) and NESCOIL $J_B$ v. free filament $J_B$ (right). The correlation is stronger in the left plot, indicating QUADCOIL is a more effective coil complexity proxy than NESCOIL.}
    \label{fig:num:param:free}
\end{figure}

Fig.\ref{fig:num:param:free} shows that a similar correlation is present between the $J_B$ values in free filament optimization and QUADCOIL. Similar to in Fig.\ref{fig:num:param:cws}, the correlation becomes stronger at high $J_B$. This correlation persists even when the solution from QUADCOIL is inexact. As Fig.\ref{fig:num:param:time} shows, the median run time of QUADCOIL is $63\times$ faster the median run time of $1398.7$s of the free filaments. This speed advantage allows QUADCOIL to be integrated into the equilibrium optimization loop. 

\begin{figure}
    \centering
    \includegraphics[width=0.5\linewidth]{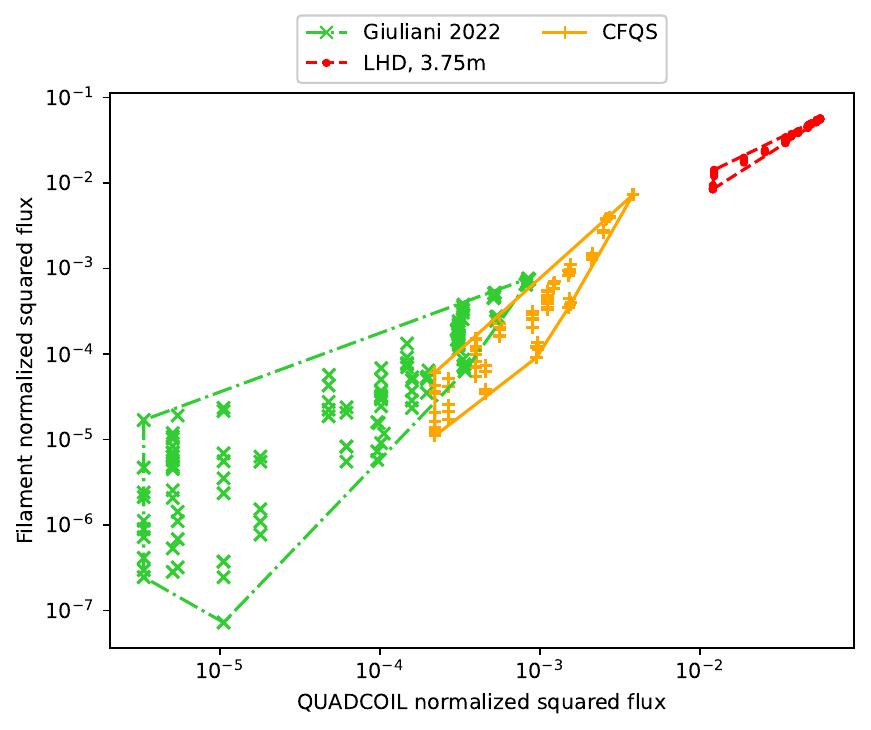}
    \caption{Comparison between the QUADCOIL and free filament $J_B$ in 3 example equilibria.}
    \label{fig:num:param:sensitivity}
\end{figure}

In addition to validating the effectiveness of QUADCOIL as a coil complexity proxy, we also find a significant difference between the best achievable $J_B$, as well as their sensitivity to $\kappa_0$, $d^\text{min}_{\text{cc}}$, and $d^\text{min}_{\text{cs}}$ across all equilibria. Fig.\ref{fig:num:param:sensitivity} compares all coil configurations in 3 example equilibria: CFQS \cite{HaifengLIU2018}, Giuliani 2022 QA \cite{giuliani_direct_2022}, and LHD major radius 3.75m \cite{iiyoshi_overview_1999}. Across all parameter combinations, the coil sets of Giuliani 2022 QA consistently achieve $J_B$ orders of magnitude lower than coils for LHD major radius 3.75m and CFQS. At the same time, coil sets for Giuliani 2022 QA are also the most sensitive to changes in $\kappa_0$, $d^\text{min}_{\text{cc}}$, and $d^\text{min}_{\text{cs}}$, while coil sets for LHD are the least. Notably, across all combinations of $\kappa_0$, $d^\text{min}_{\text{cc}}$, and $d^\text{min}_{\text{cs}}$, coil sets for each equilibrium tend to form visible "clusters". These observations indicate that the choice of equilibrium can make orders of magnitude impact on the engineering complexity of a stellarator's coils. This is potentially strong motivation for the recent uptick in single-stage optimizations. Appendix \ref{appendix:example} contains example coil sets for all three example equilibria. We will investigate the correlation between equilibrium properties and coil complexity in future studies.

\section{Conclusion}
In this paper, we presented QUADCOIL, a global winding surface method for stellarator optimization. We demonstrated its potential as an initial state tool and as a coil complexity proxy in the equilibrium stage. QUADCOIL hints at a deeper connection between filament and winding surface optimization and enables new parameter space studies that can help understand the relationship between equilibrium properties and coil complexity. We have also shown that different equilibrium choices can have orders of magnitude differences in coil feasibility. In future work, we will experiment with new objectives and constraints, such as local magnetic field errors, maximum dipole density, magnetic energy, and Lorentz force \cite{robin_minimization_2022}. We will implement new representations for $\Phi'$. A finite element basis, for example, is better suited for modeling regions with zero current. We will study the smoothness of coil complexity proxies based on QUADCOIL. By running QUADCOIL over equilibrium datasets \cite{landreman_mapping_2022, gaur_exploring_2023, giuliani_quasr:_2024}, we may identify the equilibrium properties that lead to simple, low-force coils or dipole arrays using data-driven methods. 

\section*{Acknowledgments}
This work is supported by the DOE and the Simons Foundation. We are grateful for the valuable discussions with Michael C. Zarnstorff, Rory Conlin, and other members of the Simons Collaboration on Hidden Symmetries and Fusion Energy. We thank Antoine Baillod and João Pedro Biu for their kind and timely support with the Simsopt code, and to John Kappel for providing the magnetic scale length dataset \cite{kappel_magnetic_data}.

\titleformat{\section}{\normalfont\large\bfseries}{Appendix \thesection}{1em}{}
\begin{appendices}
% \appendix
\section{Curvature proxy $\KK$}\label{appendix:KK}
In this appendix, we discuss the construction of \ref{equation:KK}, a proxy for the maximum curvature of a coil set. $\KK$ is tied to the curvature of surface current $\K$:
\begin{equation}
    \kappa = \frac{\|\KK\times\K\|_2}{\|\K\|_2^3},
\end{equation}
the variation of a surface current along each current path:
\begin{equation}
    \frac{ \partial_{\hat K} \|\K\|_2^2 }{2} = \frac{\|\KK\cdot\K\|^2_2}{\|\K\|_2^2},
\end{equation}
as well as the curvature of a filament coil with current vector $\mathbf{I}$:
\begin{equation}
    \kappa = \frac{\|\mathbf{I}\cdot\nabla\mathbf{I}\|_2}{I^2}.
\end{equation}
In practice, it's often advantageous to limit the maximum, rather than integral of an engineering metric. While $\max_{\zeta', \theta'}\|\KK\|_2$ is quartic, it is bounded by $f^\infty_\kappa$ via the norm equivalence principle.
\begin{equation}
    f^\infty_\kappa\leq\max_{\zeta', \theta'}\|\KK\|_2\leq\sqrt{3}f^\infty_\kappa.
\end{equation}
This statement is straightforward to prove. Consider a finite vector field $\vv(\rr)\in\Rset^3$ defined in an arbitrary, closed domain $\mathbf r\in\Dset$. Define:
\begin{equation*}
    \begin{split}
        \vv_2&=\argmax_\Dset(\|\vv\|_2),\\
        \vv_\infty&=\argmax_\Dset(\|\vv\|_\infty).
    \end{split}
\end{equation*}
Then, $\forall\vv\in\Dset$,
\begin{equation}
    \begin{split}
        \|\vv_2\|_2 &\geq\|\vv\|_2 \\
        \|\vv_\infty\|_\infty &\geq\|\vv\|_\infty.
    \end{split}
\end{equation}
Therefore,
\begin{equation}
    \begin{split}
        \|\vv_2\|_2&\geq\|\vv_\infty\|_2,\\
        \|\vv_2\|_\infty&\leq\|\vv_\infty\|_\infty.
    \end{split}
\end{equation}
By norm equivalence,
\begin{equation}
    \begin{split}
        \|\vv_2\|_\infty\leq&\|\vv_2\|_2\leq\sqrt{3}\|\vv_2\|_\infty,\\
        \|\vv_\infty\|_\infty\leq&\|\vv_\infty\|_2\leq\sqrt{3}\|\vv_\infty\|_\infty.\\
    \end{split}
\end{equation}
Therefore, 
\begin{equation}
    \|\vv_\infty\|_\infty\leq\|\vv_2\|_2\leq\sqrt{3}\|\vv_\infty\|_\infty,
\end{equation}
Although the functional form of $f^\infty_\kappa$ is different from the maximum curvature of a space curve, results in Section \ref{section:numerical:parameter space} have demonstrated its effectiveness.

\section{Full expressions for $\K$ and $\KK$}\label{appendix:KK expr}
This appendix derives the full expression for $\K$ and $\KK$ in term of the single-valued current potential $\Phi'$, the net poloidal current $G$, and the net toroidal current $I$, for use in the implementation.

The expression of $\K$ in term of $\Phi'$ is:
\begin{equation}
    \begin{split}
        \K &= \frac{\N}{|\N|}\times\nabla \left[\Phi'+\frac{G\zeta'}{2\pi}+\frac{I\theta'}{2\pi}\right] \\
        &= \frac{1}{|\N|}[(\partial_{\theta'}\rr)\partial_{\zeta'}-(\partial_{\zeta'}\rr)\partial_{\theta'}]\left[\Phi'+\frac{G\zeta'}{2\pi}+\frac{I\theta'}{2\pi}\right],
    \end{split}
\end{equation}
and
\begin{equation}\label{equation:K dot nabla K}
    \begin{split}
        \KK &= \K\cdot\left(\nabla\theta'\partial_{\theta'}+\nabla\zeta'\partial_{\zeta'}\right)\K\\
        &=\frac{1}{|\N|}[(\partial_{\theta'}\rr)\partial_{\zeta'}-(\partial_{\zeta'}\rr)\partial_{\theta'}]\left[\Phi'+\frac{G\zeta'}{2\pi}+\frac{I\theta'}{2\pi}\right]\cdot\left(\nabla\theta'\partial_{\theta'}+\nabla\zeta'\partial_{\zeta'}\right)\K\\
        &= \frac{1}{|\N|}[(\partial_{\zeta'}\Phi'+\frac{G}{2\pi})(\partial_{\theta'}\K)
        -(\partial_{\theta'}\Phi'+\frac{I}{2\pi})(\partial_{\zeta'}\K)].\\
    \end{split}
\end{equation}
The full expressions for $\partial_{\zeta'}\K$ and $\partial_{\theta'}\K$ are: (abbreviating $\zeta'$ or $\theta'$ as $\alpha$)
\begin{equation}
    \begin{split}
        \partial_\alpha\K =& 
        \partial_\alpha
        \left[
            \frac{\partial_{\theta'}\rr}{|\N|}(\partial_{\zeta'}\Phi'_{tot})
            -\frac{\partial_{\zeta'}\rr}{|\N|}(\partial_{\theta'}\Phi'_{tot})
        \right]\\
        =&
        \left(
            \partial_\alpha\frac{\partial_{\theta'}\rr}{|\N|}
        \right)(\partial_{\zeta'}\Phi'_{tot})
        +\frac{\partial_{\theta'}\rr}{|\N|}(\partial_{\alpha\zeta'}\Phi'_{tot})\\
        -&\left(
            \partial_\alpha\frac{\partial_{\zeta'}\rr}{|\N|}
        \right)(\partial_{\theta'}\Phi'_{tot})
        -\frac{\partial_{\zeta'}\rr}{|\N|}(\partial_{\alpha\theta'}\Phi'_{tot})\\
        =&
        \left(
            \partial_\alpha\frac{\partial_{\theta'}\rr}{|\N|}
        \right)\left(\partial_{\zeta'}\Phi'+\frac{G}{2\pi}\right)
        +\frac{\partial_{\theta'}\rr}{|\N|}(\partial_{\alpha\zeta'}\Phi')\\
        -&\left(
            \partial_\alpha\frac{\partial_{\zeta'}\rr}{|\N|}
        \right)\left(\partial_{\theta'}\Phi'+\frac{I}{2\pi}\right)
        -\frac{\partial_{\zeta'}\rr}{|\N|}(\partial_{\alpha\theta'}\Phi')
    \end{split}
\end{equation}
The full expression for $\partial_\alpha\frac{\partial_{\zeta'}\rr}{|\N|}$ and $\partial_\alpha\frac{\partial_{\theta'}\rr}{|\N|}$ are: (abbreviating $\zeta'$ or $\theta'$ as $\beta$)
\begin{equation*}
    \begin{split}
         \partial_\alpha\frac{\partial_\beta\rr}{|\N|} 
         &= \frac{\partial_{\alpha\beta}\rr}{|\N|} 
         + (\partial_\beta\rr)\partial_\alpha\left(\frac{1}{|\N|}\right) \\
         &= \frac{\partial_{\alpha\beta}\rr}{|\N|} 
         + (\partial_\beta\rr)\partial_\alpha(\N\cdot\N)^{-1/2} \\
         &= \frac{\partial_{\alpha\beta}\rr}{|\N|} 
         + (\partial_\beta\rr)\partial_\alpha(\N\cdot\N)\left(-\frac{1}{2}\right)(\N\cdot\N)^{-3/2} \\
         &= \frac{\partial_{\alpha\beta}\rr}{|\N|} 
         - (\partial_\beta\rr)\frac{\partial_\alpha(\N\cdot\N)}{2|\N|^3}.
    \end{split}
\end{equation*}
Here, $\N\equiv\frac{\partial\rr}{\partial\zeta'}\times\frac{\partial\rr}{\partial\theta'}$, $\partial_\alpha\N=\frac{\partial^2\rr}{\partial\alpha\partial\zeta'}\times\frac{\partial\rr}{\partial\theta'} + \frac{\partial\rr}{\partial\zeta'}\times\frac{\partial^2\rr}{\partial\alpha\partial\theta'}$, $\partial_\alpha(\N\cdot\N) = 2\N\cdot\partial_\alpha\N$. 

\section{Relaxation procedures}
\label{appendix:shor procedures}
In this appendix, we find a convex relaxation to \eqref{equation:num:KK original} following a procedure common in quadratic optimization, known as Shor Relaxation \cite{shor_quadratic_1987}. This procedure converts the problem into a minimization problem for a linear objective on a non-convex domain, and then solves the problem in a convex superset of the domain. 

Consider a scalar quadratic function of $y\equiv(\Phi', a)\in\Rset^{n_{\Phi'}+n_a}$:
\begin{gather}
    f(\Phi', a) = y^T Q y + \alpha^T y + \beta, \\ \notag
    Q\in\Rset^{(n_{\Phi'}+n_a)\times (n_{\Phi'}+n_a)}, \alpha\in\Rset^{(n_{\Phi'}+n_a)}, \beta\in\Rset.
\end{gather}
To make $f$ linear, we homogenize $f\rightarrow f_h$ by introducing an additional scalar degree of freedom, $z$:
\begin{align}
f_h(\Phi', a,z) = y^T Q y + \alpha^T y z + \beta z^2.
\end{align}
The new function satisfies $f_h(\Phi', a, 1) = f(\Phi', a)$ and is bilinear in $x\equiv(\Phi', a, z)\in\Rset^{(n_{\Phi'}+n_a+1)}$. For simplicity, define $n_x\equiv(n_{\Phi'}+n_a+1)$. Introduce a matrix form of the unknown, $X\equiv xx^T$, then $f$ is linear in $X$:
\begin{equation}
\label{equation:homogenize}
    \begin{split}
        f_h(\Phi', a,z) 
        &=  y^T Q  y + \alpha^T  y z + \beta c = \text{tr}(FX)\\
        &= \text{tr}\left[
            \begin{pmatrix}
                Q & \alpha \\ 
                0 & \beta \\
            \end{pmatrix}
            \begin{pmatrix}
                & y_1 y_1 &... &  y_1 y_{(n_x-1)} &  y_1z\\
                &\vdots &\ddots & \vdots & \vdots\\
                & y_{(n_x-1)} y_1 &... &  y_{(n_x-1)} y_{(n_x-1)} &  y_nz\\
                & y_1z &... &  y_{(n_x-1)}z &z^2\\
            \end{pmatrix}
        \right]\\
        &\text{or, equivalently, }\\
        &= \text{tr}\left[
            \begin{pmatrix}
                \frac{1}{2}Q + \frac{1}{2}Q^T & \frac{1}{2}\alpha\\
                \frac{1}{2}\alpha^T & \beta \\
            \end{pmatrix}
            \begin{pmatrix}
                & y_1 y_1 &... &  y_1 y_{(n_x-1)} &  y_1z\\
                &\vdots &\ddots & \vdots & \vdots\\
                & y_{(n_x-1)} y_1 &... &  y_n y_{(n_x-1)} &  y_nz\\
                & y_1z &... &  y_{(n_x-1)}z &z^2\\
            \end{pmatrix}
        \right]\\
        &(\text{some solvers work better with symmetric matrices}).\\
    \end{split}
\end{equation}
Following this procedure, we can rewrite $f, g_j, h_k$ from the QCQP in Eq.~\eqref{equation:problem:raw} into $\text{tr}(FX), \text{tr}(G_jX), \text{tr}(H_kX)$, and transform the equation into a linear program (LP) on a closed, non-convex, compact domain $\Dset$:
\begin{equation}
\label{equation:problem:LP}
        \min_{X\in\Dset, b\in\Rset^{n_b}}\left[\text{tr}(FX) +c^Tb\right].
\end{equation}
Here, $\Dset$ is the intersection between a polyhedron $\Dset_p$ and a non-convex set $\Dset_n$:
\begin{gather}
    \Dset = \Dset_p\cap\Dset_n\\
    \begin{split}
        \Dset_p = \{X|X\in\Rset^{n_x^2}, 
        &\text{tr}(G_jX)+d_j^Tb\leq e_j, \\
        & \text{tr}(H_kX) + m_k^Tb = n_k, 
        \}
    \end{split}\\
    \begin{split}
        \Dset_n &= \{xx^T|x\in\Rset^{n_x}, x_{n_x}=1\}\\
        &= \{X|X\in \Sset^{n_x}_+,X_{n_x, n_x}=1, \rank(X)=1\}.
    \end{split}
\end{gather}
Here, $\Sset_+^{n_x}$ denotes the set of $n_x\times n_x$ positive semi-definite matrices. The polyhedron $\Dset_p$ contains the information of all quadratic constraints, and $\Dset_n$ restricts the form of the new unknown $X$ so that the homogenized problem reduces to the original problem. Equation~\eqref{equation:problem:LP} is equivalent to an LP on the convex hull of its domain:
\begin{equation}
\label{equation:problem:LP hull}
        \min_{X\in \conv(\Dset), b\in\Rset^{n_b}}\left[\text{tr}(FX) + c^Tb\right].
\end{equation}
For most choice of $f$ and $G_j$, $\conv(\Dset)$ does not have a simple expression. However, by the non-decreasing property of convex hull, $\conv(\Dset)\subseteq\conv(\Dset_p)\cap \conv(\Dset_n)$. Because $\Dset$ is a polyhedron, $\conv(\Dset_p)=\Dset_p$. We will prove at the end of this appendix, that $\conv(\Dset_n)$ is simply: 
\begin{equation}
\label{equation:convex hull}
    \conv(\Dset_n) = \{X|X\in \Sset^{n_x}_+,X_{n_x, n_x}=1\}.
\end{equation}
The change of domain completes the convex relaxation of \eqref{equation:problem:raw}, as:
\begin{equation}
\label{equation:problem:conic}
    \begin{split}
        \min&\left[\text{tr}(FX) +c^Tb\right],\\
        X&\in\Dset_p\cap \conv(\Dset_n),\\
        b&\in\Rset^{n_b}.
    \end{split}
\end{equation}

\begin{figure}
    \centering
    \includegraphics[width=0.6\linewidth]{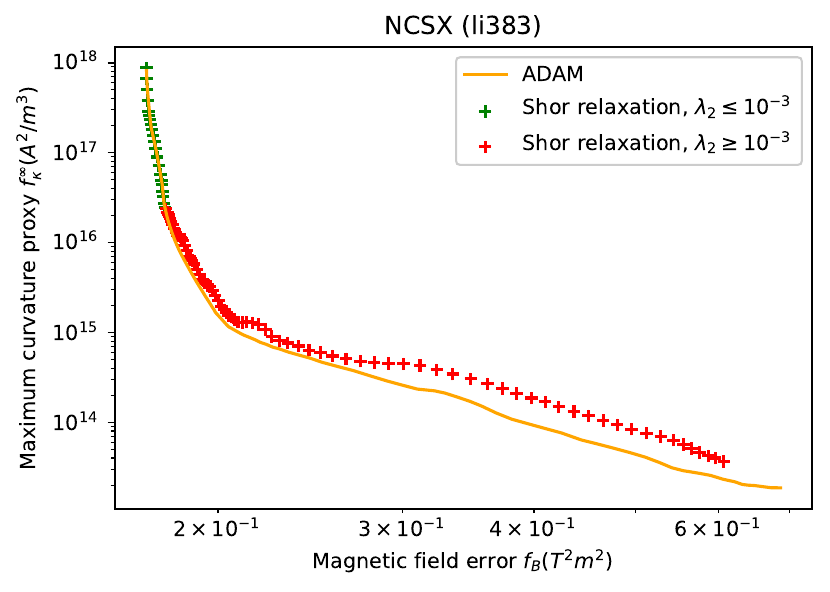}
    \caption{The numerical signature for exact Shor relaxation. The green dots indicate cases where the numerical exactness check, $|\lambda_2/\lambda_1|\leq \lambda_\text{crit}$, succeeded, and the red dot indicates cases where the check fails. From the left to the right, $f_B$ increases, the strength of the convex penalty decreases, and eventually, Shor relaxation breaks. The transition from green to red coincides with the point where ADAM starts to outperform Shor relaxation. This validates the effectiveness of the numerical exactness check.}
    \label{fig:appendix:validity}
\end{figure}

It is easy to see that the solution $X$ exactly solves \eqref{equation:problem:raw} when $\rank(X)=1$. In practice, this gives a fast numerical test for the optimality of a Shor relaxation solution. Then, we can test its optimality (or whether $\rank(X)=1$) by comparing the ratio of the leading eigenvalue $\lambda_1$ of $X$ with the second largest eigenvalue, $\lambda_2$, with a small, empirical threshold $\lambda_\text{crit}$:
\begin{equation}
\label{equation:appendix:test}
    |\lambda_2/\lambda_1|\leq \lambda_\text{crit}.
\end{equation}
In this paper, we choose $\lambda_\text{crit}=10^{-3}$. Fig.~\ref{fig:appendix:validity} compares the results solving~\eqref{equation:problem:raw} with Shor relaxation and the stochastic, non-convex optimization algorithm, ADAM, from all three cases in Section~\ref{section:numerical:KK}. Note that when $|\lambda_2|$ becomes greater than $10^{-3}$, ADAM begins to outperform Shor relaxation, indicating that the relaxation no longer yields an exact solution to \eqref{equation:problem:raw}. 
Now, we return to prove that the convex hull of $\Dset_n$,
\begin{equation}
    \conv(\Dset_n) 
    = \left\{\left.
        \sum_{i\in \{1,2,...\}} a_iX_i
        \right|
        a_i\in[0,1], \sum_i^{n_{\text{obj}}}a_i=1, 
        X_i\in \Dset_n
    \right\},
\end{equation}
is:
\begin{equation}
    \Hset_n\equiv \{X|X\in \Sset^{n_x}_+, X_{n_x, n_x}=1\}.
\end{equation}
To start, we first prove $\conv(\Dset_n)\subseteq \Hset_n$. Substitute in the definition of $\Dset_n = \{xx^T|x\in\Rset^{n_x}, x_{n_x}=1\}$:
\begin{equation}
    \conv(\Dset_n) 
    = \left\{\left.
            \sum_{i\in \{1,2,...\}} a_ix_ix_i^T
            \right|
            a_i\in[0,1], \sum_i^{n_{\text{obj}}}a_i=1, 
            x_i\in\Rset^{n_x},
            (x_i)_{n_x}=1
        \right\},
\end{equation}
Then, because the last element in $x_i$, $(x_i)_{n_x}=1$, the last diagonal element of $x_ix_i^T$, $(x_ix_i^T)_{n_x, n_x}=1$. This allows us to re-write the summation, $\sum_i^{n_{\text{obj}}}a_i=1$, as:
\begin{equation}\label{equation:appendix:Th}
        \conv(\Dset_n) = \left\{\left.
            \sum_{i\in \{1,2,...\}} a_ix_ix_i^T
            \right|
            a_i\in[0,1], \sum_i^{n_{\text{obj}}}a_i(x_ix_i^T)_{n_x, n_x}=1, 
            x_i\in\Rset^{n_x},
            (x_i)_{n_x}=1
        \right\}
\end{equation}
Because the sum of positive semi-definite matrices is still positive semi-definite, 
\begin{equation}
    \sum_{i\in \{1,2,...\}} a_ix_ix_i^T
\end{equation}
is positive semi-definite. It is easy to see, then, that:
\begin{equation}
    \conv(\Dset_n)\subseteq\Hset_n\equiv \{X|X\in \Sset^{n_x}_+, X_{n_x, n_x}=1\}.
\end{equation}
This completes half of the proof. 

To prove $\Hset_n\subseteq \conv(\Dset_n)$, we consider the rank-1 decompositions of $X$. It is easy to see that $X$ has an infinite number of orthogonal, rank-1 decompositions. Starting from the eigendecomposition of $X$;
\begin{equation}
    X = Q_+\Lambda_+Q_+^T.
\end{equation}
Here, $\Lambda_+$ is a $\rank(X)$, positive diagonal matrix that contains all non-zero eigenvalues of $X$, and $Q$ is a $\rank(X)\times n_x$ matrix containing $\rank(X)$ orthonormal, real eigenvectors of $X$. This decomposition is non-unique to an arbitrary orthogonal transformation $R$, allowing us to write an infinite number of rank-1 decompositions as orthogonal vectors $m_i, i = 1, ..., \rank(X)$:
\begin{equation}
    X = \sum_{i=1}^{\rank(X)}m_im_i^T = MM^T.
\end{equation}
Here, $m_i$ are the rows of $M\equiv R\sqrt\Lambda_+ Q_+\in\Rset^{\rank(X)\times n_x}$. $m_i$ are orthogonal, but not normal. 

Now, consider $(\sqrt\Lambda_+ Q_+)_{n_x}$, the last column of $\sqrt\Lambda_+ Q_+$. Because $X_{n_x, n_x}=1$, $|(\sqrt\Lambda_+ Q_+)_{n_x}|^2=1$. Because orthogonal transforms are length-preserving, $(m_i)_{n_x}$ must also satisfy $\sum_{i=1}^{\rank(X)}(m_i)^2_{n_x}=1$. Choose an $R$ that produces non-zero $(m_i)_{n_x}$ for all $i$. This choice of $R$ will define a rank-1 decomposition satisfying:
\begin{equation}
     X = \sum_{i=1}^{\rank(X)}m_im_i^T, \sum_{i=1}^{\rank(X)}(m_i)^2_{n_x}=1, 
\end{equation}
Because $\sum_{i=1}^{\rank(X)}(m_i)^2_{n_x}=1$, and $m_i$ are real, $(m_i)^2_{n_x}\in(0, 1]$. Define $a_i\equiv(m_i)^2_{n_x}$, then:
\begin{equation}
\label{equation:appendix:hull:last}
\begin{split}
    X &= \sum_{i=1}^{\rank(X)}a_i\left[\frac{m_i}{(m_i)}\right]\left[\frac{m_i}{(m_i)}\right]^T, a_i\in(0,1], \sum_{i=1}^{\rank(X)}a_i=1\\
    &\in \left\{\left.
            \sum_{i\in \{1,2,...\}} a_ix_ix_i^T
            \right|
            a_i\in[0,1], \sum_i^{n_{\text{obj}}}a_i=1, 
            x_i\in\Rset^{n_x},
            (x_i)_{n_x}=1
        \right\}\\
    &\in\conv(\Dset_n).
\end{split}
\end{equation}
This proves $\Hset_n\subseteq \conv(\Dset_n)$. It may be tempting to apply \eqref{equation:appendix:hull:last} directly to the eigendecomposition of $X$. However, because there is no guarantee that the last elements of all eigenvectors are non-zero, the use of $R$ and $M$ is necessary.

By proving $\conv(\Dset_n)\subseteq\Hset_n $ and  $\Hset_n\subseteq \conv(\Dset_n)$, we have proved that $\conv(\Dset_n)=\Hset_n=\{X|X\in \Sset_+^{n_x}, X_{n_x,n_x}=1\}$. 

\section{Random samples of free filament topology}
\label{appendix:example}
\begin{figure}
    \centering
    \includegraphics[width=0.8\linewidth]{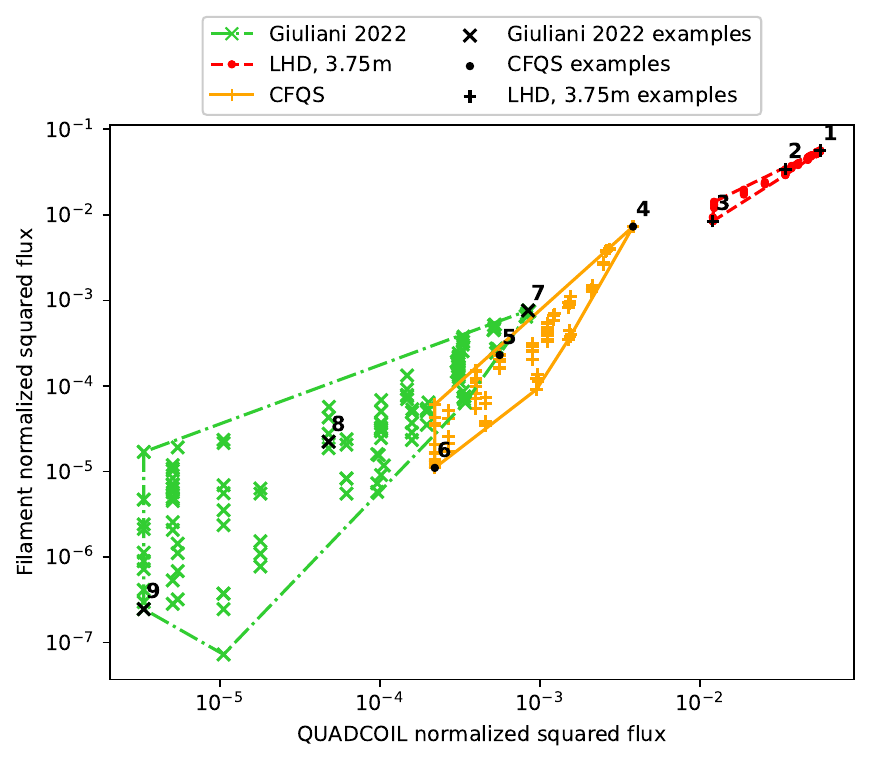}
    \caption{The locations of the 9 example cases. Case 1-3 are optimized for the LHD 3.75m equilibrium, case 4-6 are optimized for the CFQS equilibrium, and case 7-9 are optimized for the Giuliani 2022 equilibrium. All cases in this figure have exact Shor relaxation and convergent free filament optimization with $\leq5\%$ constraint breaking.}
    \label{fig:appendix:example:loc}
\end{figure}

\begin{figure}
    \centering
    \begin{tabular}{ccc}
        \begin{subfigure}[b]{0.3\textwidth}
            \centering
            \includegraphics[width=\textwidth]{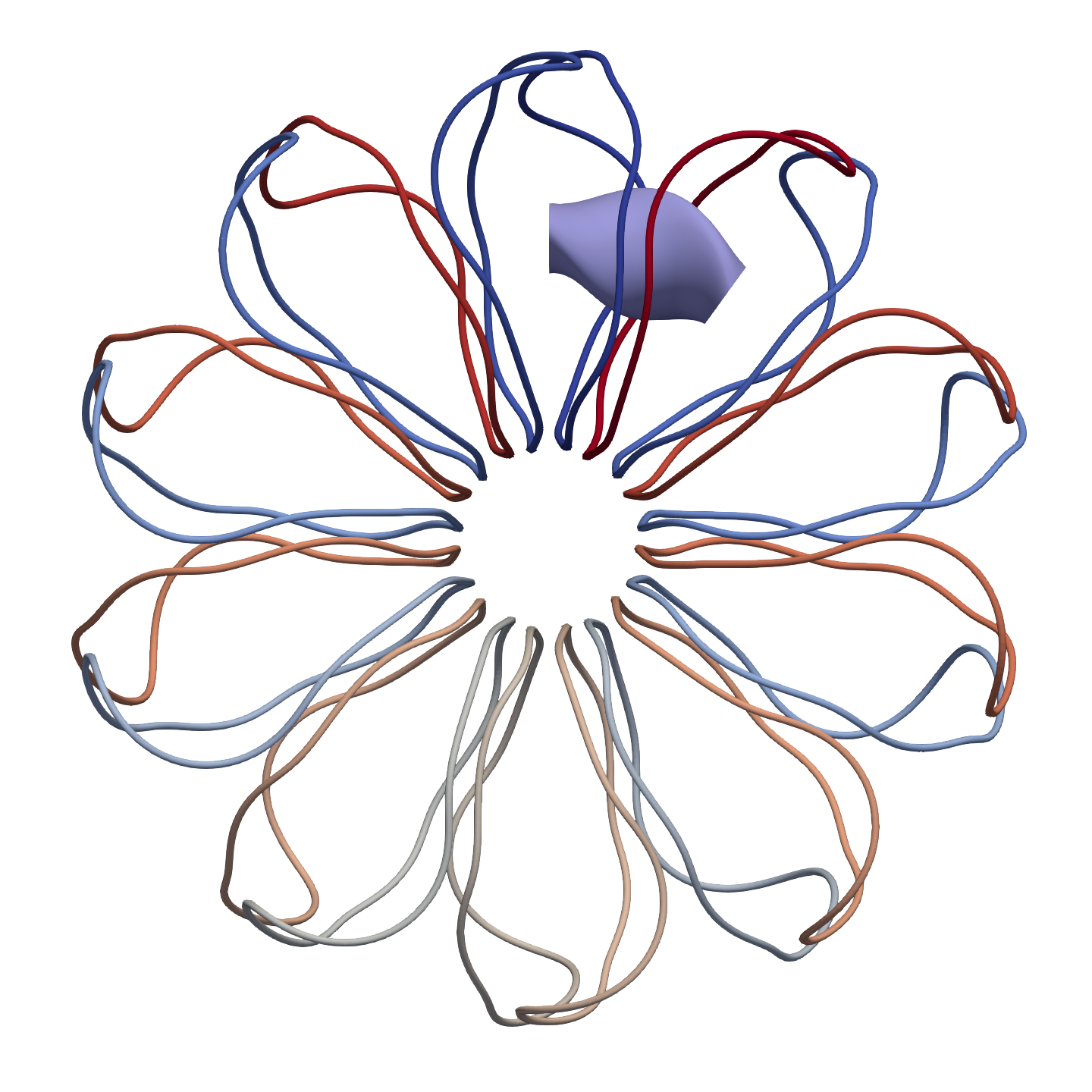}
            \caption{Case no.1}
        \end{subfigure} &
        \begin{subfigure}[b]{0.3\textwidth}
            \centering
            \includegraphics[width=\textwidth]{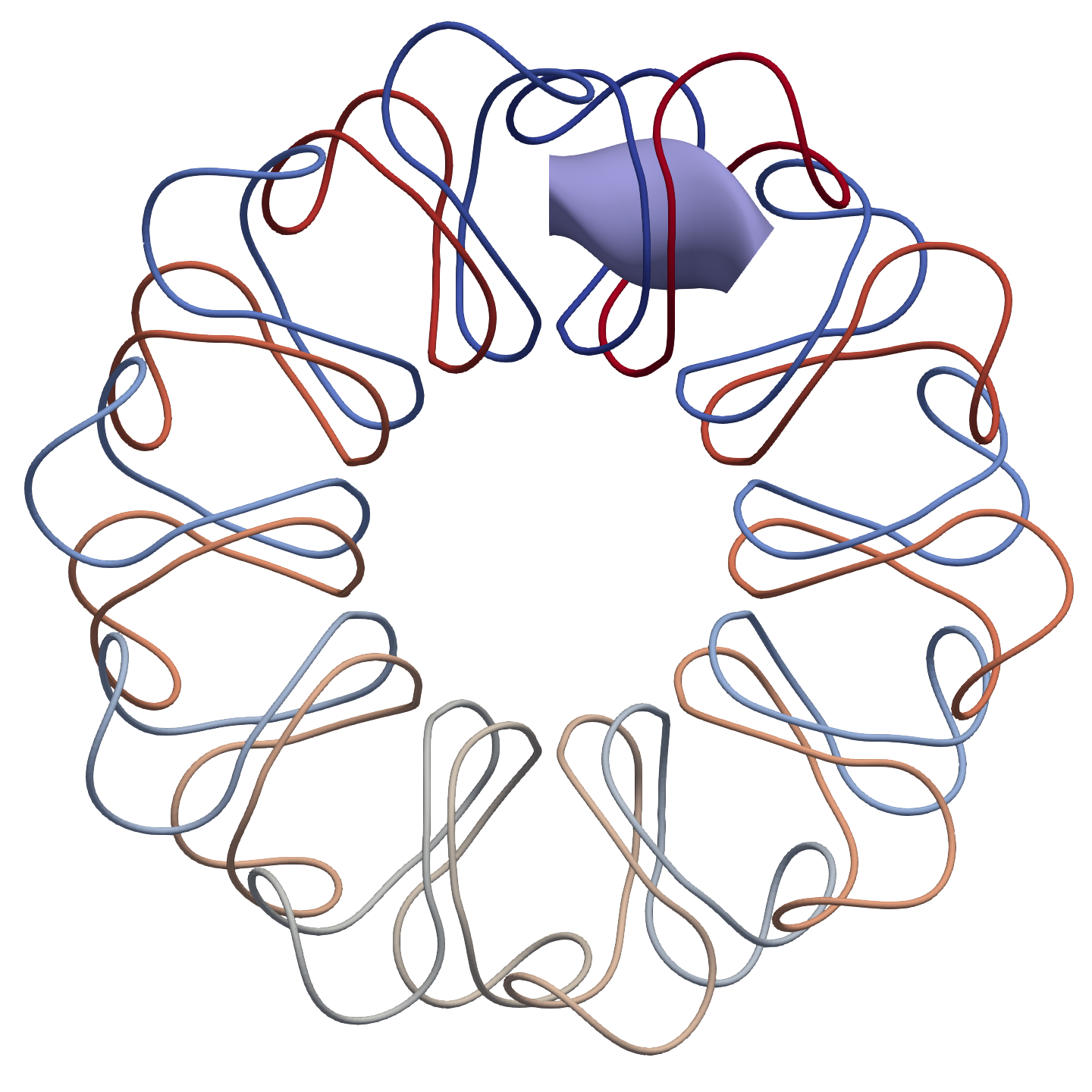}
            \caption{Case no.2}
        \end{subfigure} &
        \begin{subfigure}[b]{0.3\textwidth}
            \centering
            \includegraphics[width=\textwidth]{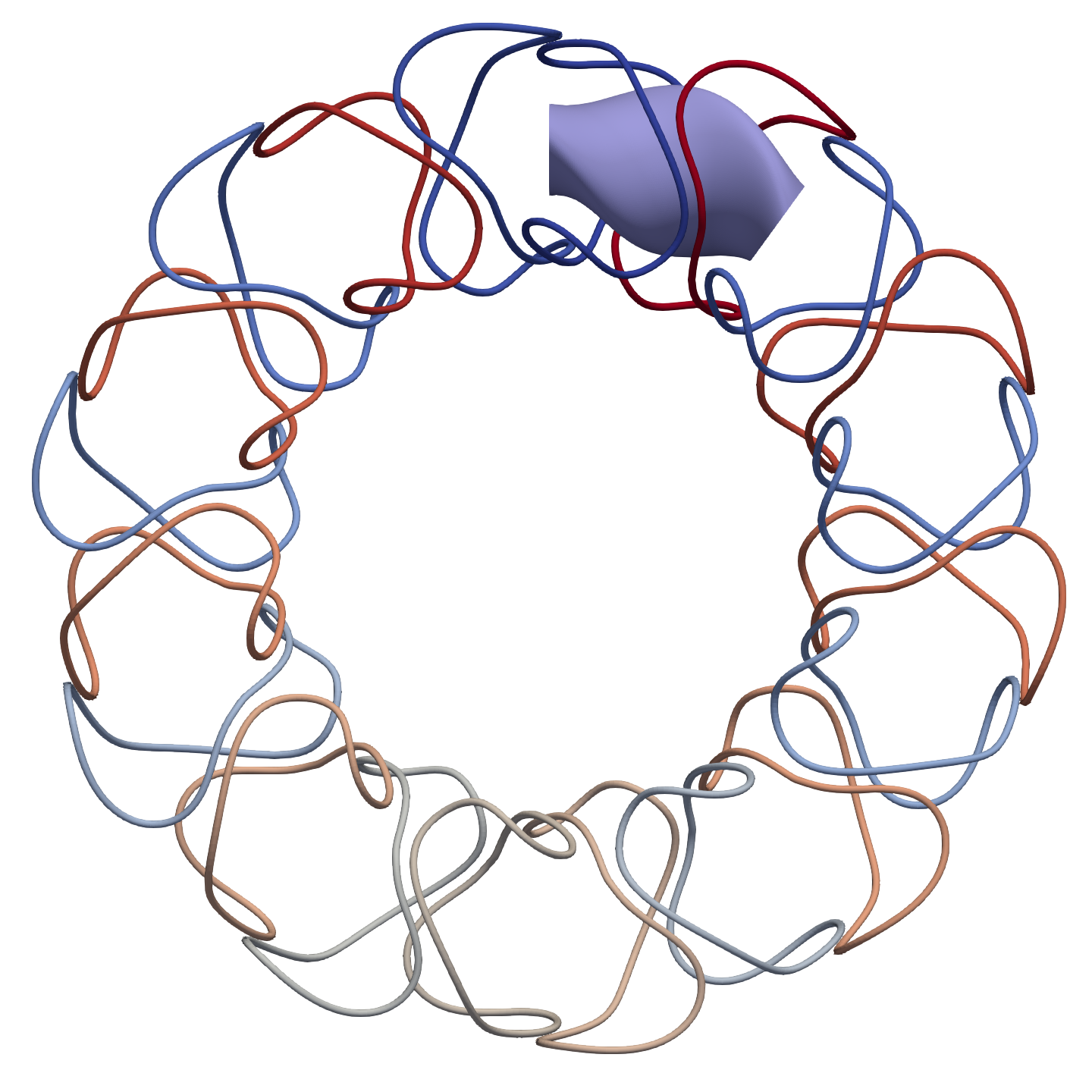}
            \caption{Case no.3}
        \end{subfigure} 
    \end{tabular}
    \caption{Example coil sets (cases 1 - 3 in \ref{fig:appendix:example:loc}) for the LHD 3.75m equilibrium. The normalized squared flux decreases from case 1 - 3.}
    \label{fig:appendix:example:LHD}
\end{figure}

\begin{figure}
    \centering
    \begin{tabular}{ccc}
        \begin{subfigure}[b]{0.3\textwidth}
            \centering
            \includegraphics[width=\textwidth]{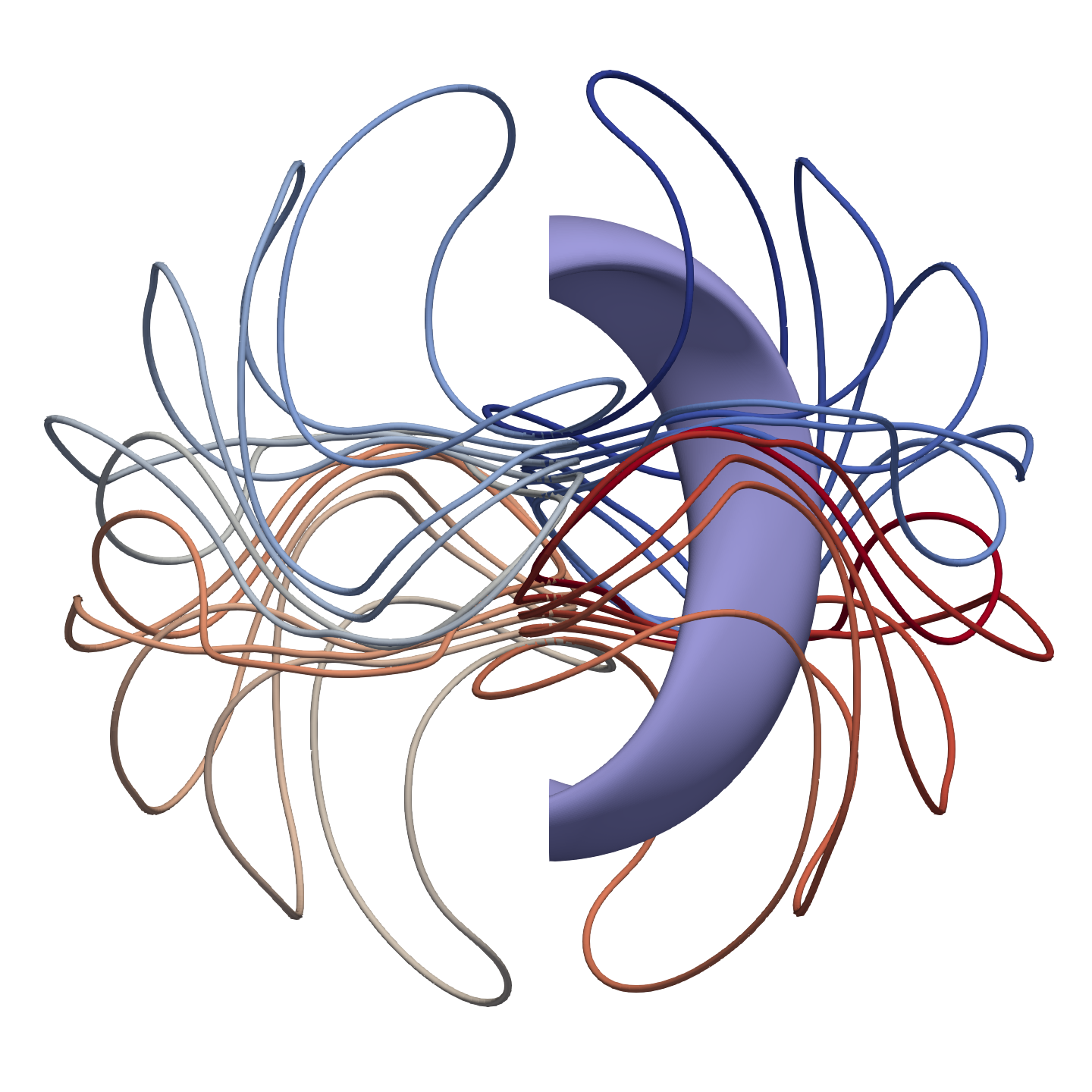}
            \caption{Case 4}
        \end{subfigure} &
        \begin{subfigure}[b]{0.3\textwidth}
            \centering
            \includegraphics[width=\textwidth]{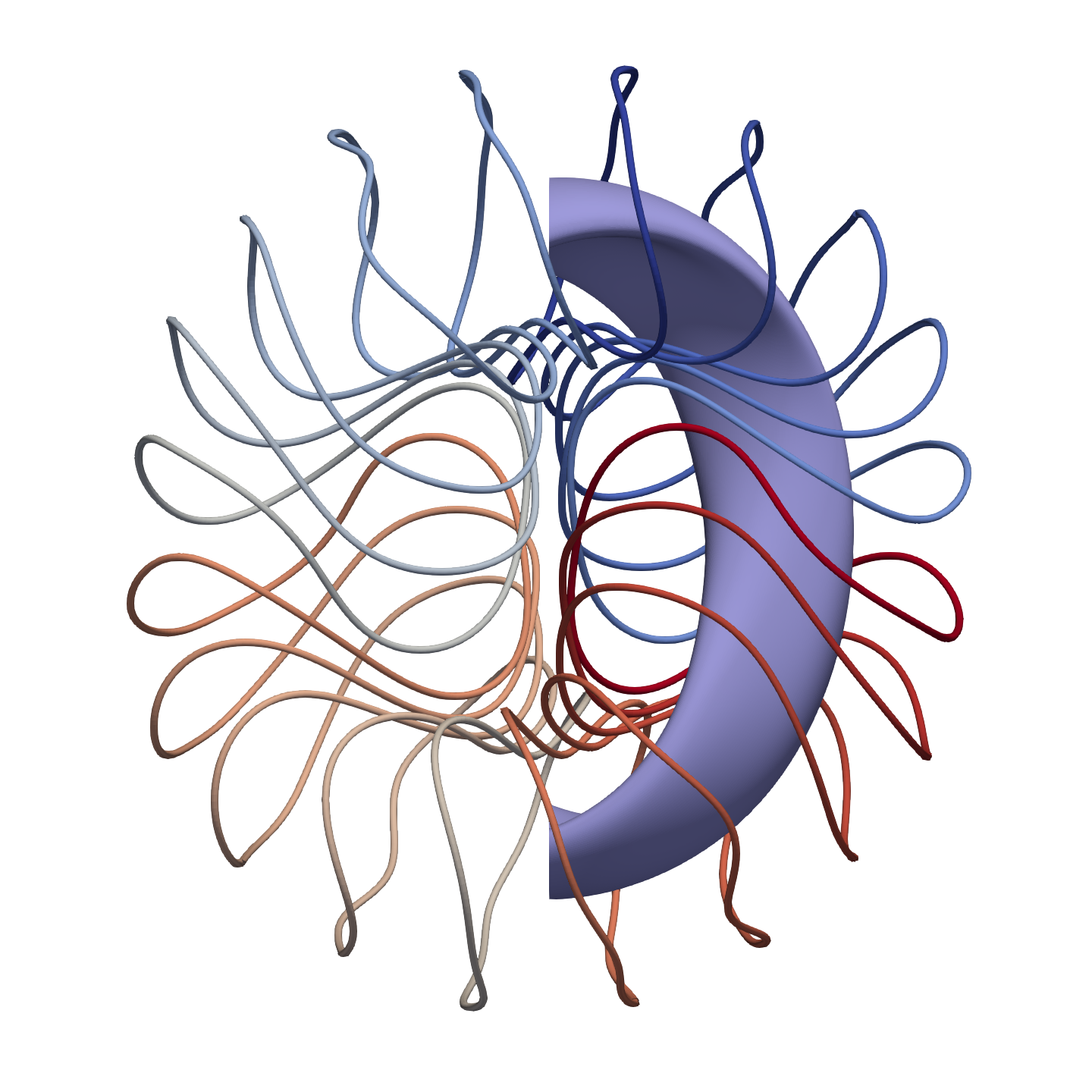}
            \caption{Case 5}
        \end{subfigure} &
        \begin{subfigure}[b]{0.3\textwidth}
            \centering
            \includegraphics[width=\textwidth]{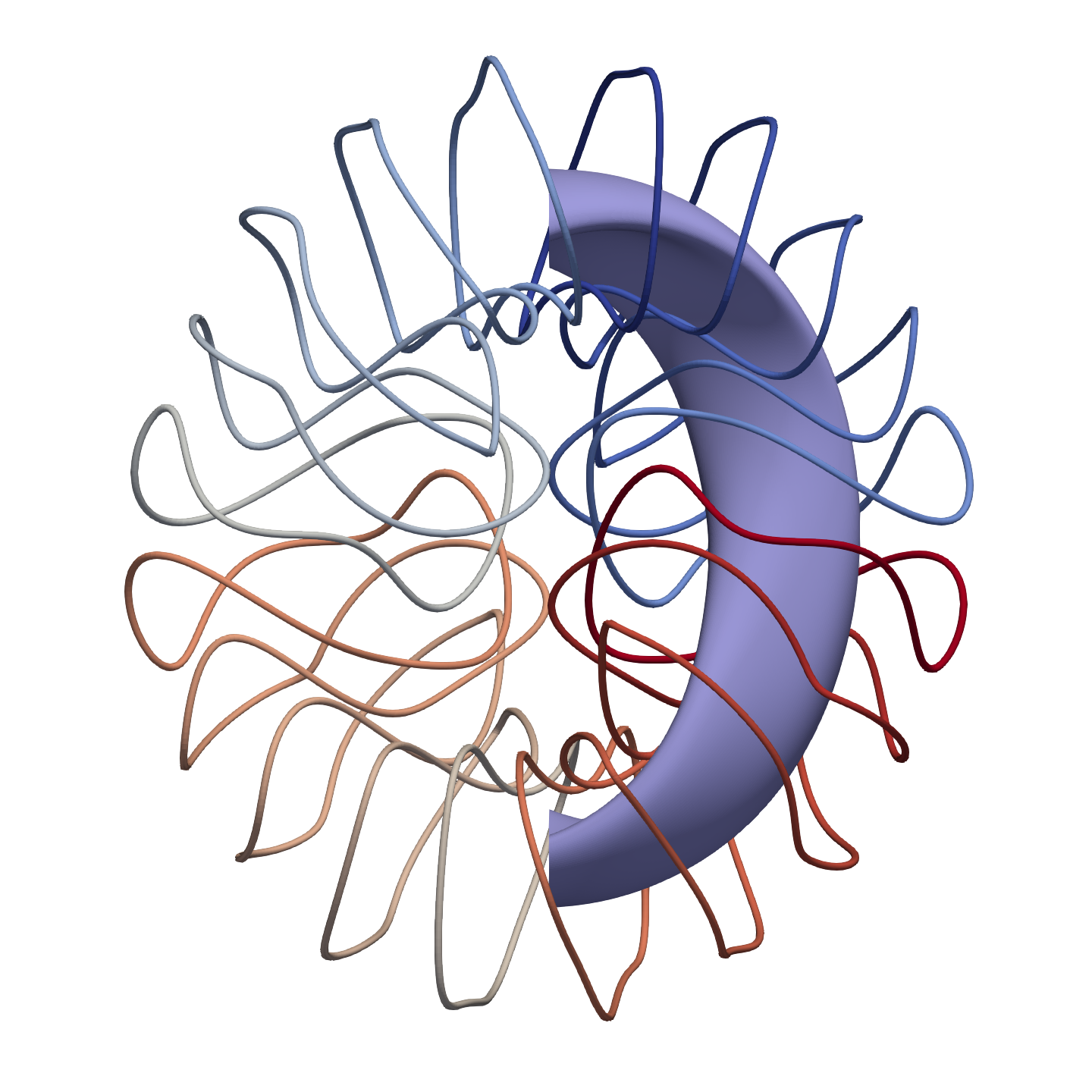}
            \caption{Case 6}
        \end{subfigure} 
    \end{tabular}
    \caption{Example coil sets (cases 4 - 6 in \ref{fig:appendix:example:loc}) for the CFQS equilibrium. The normalized squared flux decreases from case 4 - 6.}
    \label{fig:appendix:example:CFQS}
\end{figure}

\begin{figure}
    \centering
    \begin{tabular}{ccc}
        \begin{subfigure}[b]{0.3\textwidth}
            \centering
            \includegraphics[width=\textwidth]{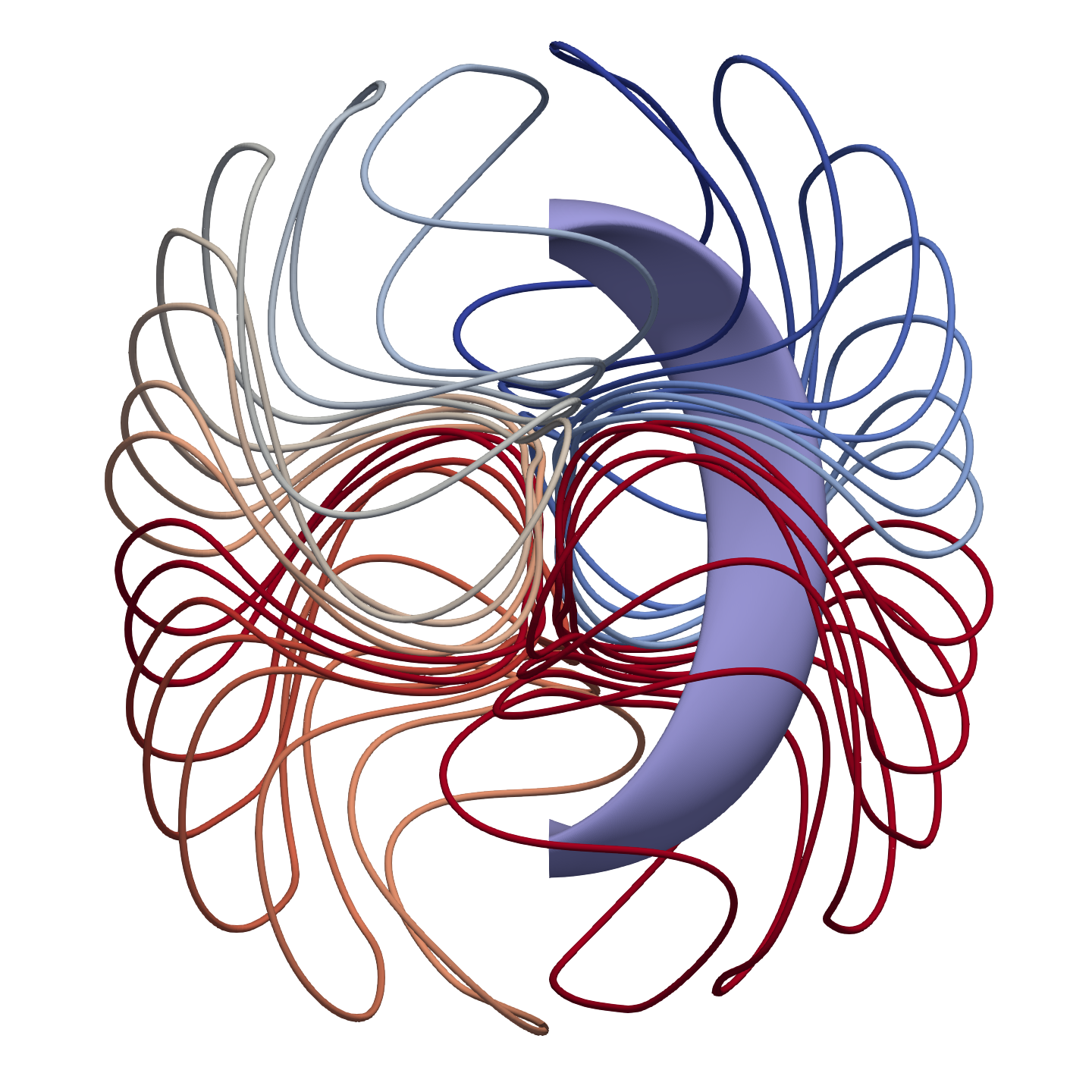}
            \caption{Case 7}
        \end{subfigure} &
        \begin{subfigure}[b]{0.3\textwidth}
            \centering
            \includegraphics[width=\textwidth]{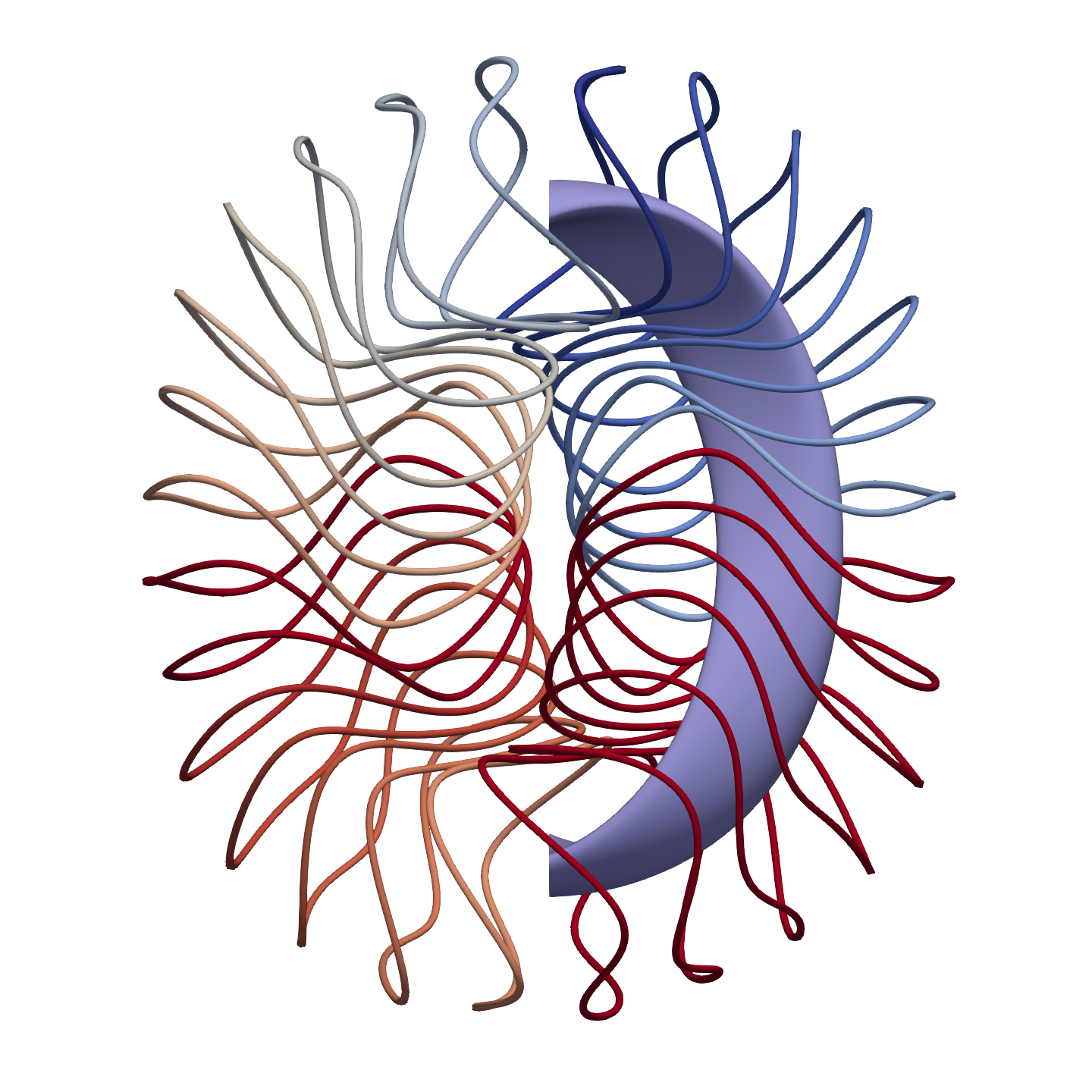}
            \caption{Case 8}
        \end{subfigure} &
        \begin{subfigure}[b]{0.3\textwidth}
            \centering
            \includegraphics[width=\textwidth]{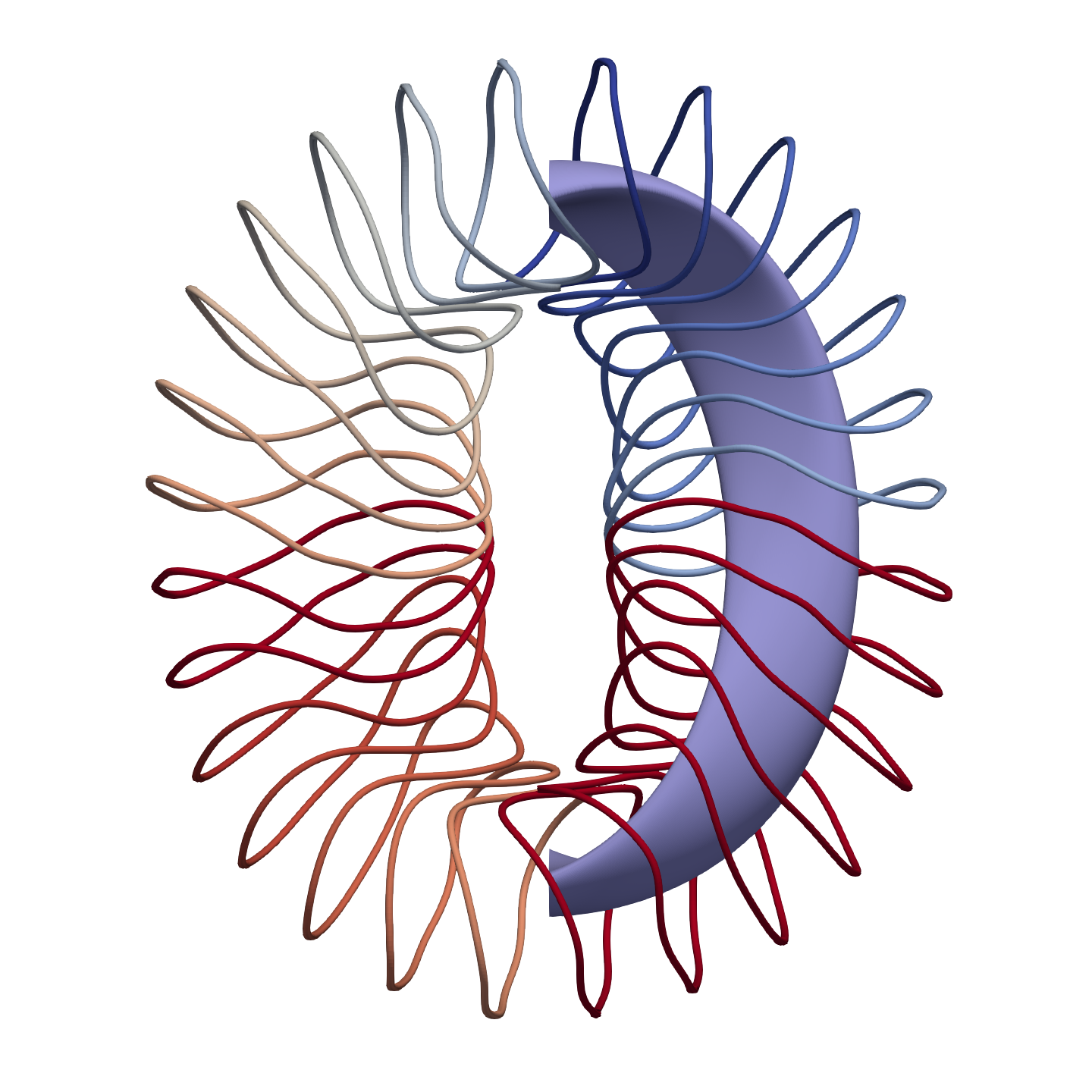}
            \caption{Case 9}
        \end{subfigure} 
    \end{tabular}
    \caption{Example coil sets (cases 7 - 9 in \ref{fig:appendix:example:loc}) for the Giuliani 2022 equilibrium. The normalized squared flux decreases from case 7 - 9.}
    \label{fig:appendix:example:giuliani}
\end{figure}

As shown in Fig. \ref{fig:appendix:example:loc}, we choose 3 examples each from the LHD, CFQS, and Giuliani 2022 equilibria with low, intermediate and high $J_B$. Fig. \ref{fig:appendix:example:LHD} through Fig. \ref{fig:appendix:example:giuliani} show the optimum free filament coils in these examples. 

%fig:appendix:example:major

\end{appendices}

\bibliographystyle{unsrt}  
\bibliography{references}

\end{document}